\documentclass[aps,prd,twocolumn,nofootinbib,superscriptaddress,showpacs,preprintnumbers,notitlepage]{revtex4-1}

\usepackage{amsmath,amsfonts,amssymb}
\usepackage{bm,color,xcolor}
\usepackage{epsfig,dcolumn}

\DeclareMathOperator{\re}{Re}
\DeclareMathOperator{\im}{Im}

\begin{document}

\title{Vector Meson Photoproduction with a Linearly Polarized Beam}

\author{V.~Mathieu}
\email{vmathieu@jlab.org}
\affiliation{Theory Center, Thomas Jefferson National Accelerator Facility, Newport News, VA 23606, USA}

\author{J.~Nys}
\affiliation{Theory Center, Thomas Jefferson National Accelerator Facility, Newport News, VA 23606, USA}
\affiliation{Department of Physics and Astronomy, Ghent University, Belgium}
\affiliation{Center for Exploration of Energy and Matter, Indiana University, Bloomington, IN 47403, USA}
\affiliation{Physics Department, Indiana University, Bloomington, IN 47405, USA}


\author{C.~Fern\'andez-Ram\'irez}
\affiliation{Instituto de Ciencias Nucleares, Universidad Nacional Aut\'onoma de M\'exico, Ciudad de M\'exico 04510, Mexico}

\author{A.~Jackura}
\affiliation{Center for Exploration of Energy and Matter, Indiana University, Bloomington, IN 47403, USA}
\affiliation{Physics Department, Indiana University, Bloomington, IN 47405, USA}


\author{A.~Pilloni}
\affiliation{Theory Center, Thomas Jefferson National Accelerator Facility, Newport News, VA 23606, USA}

\author{N.~Sherrill}
\affiliation{Center for Exploration of Energy and Matter, Indiana University, Bloomington, IN 47403, USA}
\affiliation{Physics Department, Indiana University, Bloomington, IN 47405, USA}

\author{A.~P.~Szczepaniak}
\affiliation{Theory Center, Thomas Jefferson National Accelerator Facility, Newport News, VA 23606, USA}
\affiliation{Center for Exploration of Energy and Matter, Indiana University, Bloomington, IN 47403, USA}
\affiliation{Physics Department, Indiana University, Bloomington, IN 47405, USA}

\author{G.~Fox}
\affiliation{School of Informatics, Computing, and Engineering, Indiana University, Bloomington, IN 47405, USA}

\collaboration{Joint Physics Analysis Center}

\preprint{JLAB-THY-18-2650}


\begin{abstract}
We propose a model based on Regge theory to describe photoproduction of light vector mesons. 
We fit the SLAC data and make predictions for the energy and momentum transfer dependence of the spin-density matrix elements in photoproduction of  $\omega$, $\rho^0$ and $\phi$  mesons at $E_\gamma \sim 8.5$ GeV, which are soon to be measured at Jefferson Lab. 
\end{abstract}

\maketitle

\section{Introduction}
With the recent development of the 12 GeV electron beam at Jefferson Lab (JLab)~\cite{Ghoul:2015ifw,battaglieri2005meson}, new  precision measurements of  light meson photoproduction and electroproduction are expected in the near future. These will provide constraints on resonance production dynamics, including production of gluonic excitations. For example, the GlueX measurement of the photon beam asymmetry in the production of $\pi^0$ and $\eta$ mesons~\cite{AlGhoul:2017nbp} established the dominance of natural-parity $t$-channel exchanges for production in the forward direction~\cite{Nys:2016vjz}. This measurement seems to contradict earlier SLAC data~\cite{Anderson:1971xh} that suggests significant contribution from unnatural-parity exchanges. It was shown in  \cite{Mathieu:2015eia} that the weak energy dependence of the axial-vector contributions suggested by the SLAC data is difficult to reconcile with predictions from Regge theory, while  the GlueX data seem to be more in line with theory predictions. The GlueX  measurement, however, was performed at fixed photon energy. Nevertheless, more data from both GlueX and CLAS12 will be needed to refine our understanding of the production mechanisms. 

\begin{figure}[htb] 
\begin{center}
\includegraphics[width=0.5\linewidth]{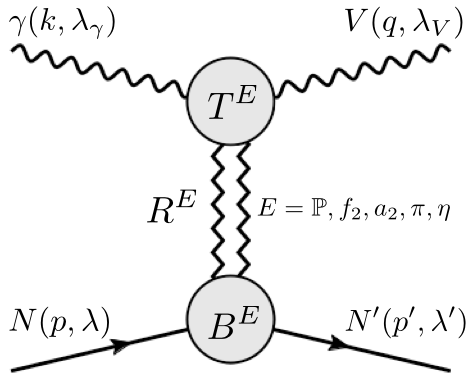}
\end{center}
\caption{\label{fig:diagram} Schematic representation of the factorized amplitude of a Regge exchange $E$ in Eq.~\eqref{eq:factor}. The photon and nucleon vertices are denoted by $T^E$ and $B^E$, respectively. The Regge propagator of the exchange $E$ is $R^E$.}
\end{figure}

We consider the reaction $\gamma(k,\lambda_\gamma) N(p,\lambda)\to V(q,\lambda_V) N'(p',\lambda')$. At high energies, the amplitude in the forward direction is dominated by exchange of Regge poles (Reggeons).
As illustrated in Fig.~\ref{fig:diagram}, the Reggeon amplitude factorizes into a product of two vertices. The upper vertex describes the beam (photon) interactions, and the lower vertex describes the target (proton) interactions. The Mandelstam variables are $s = (k+p)^2$ and $t = (k-q)^2$.
  Factorization of Regge vertices follows from unitarity in the $t$-channel, where Regge pole is a common pole in all partial waves related by unitarity and its vertices determine residues of the poles~\cite{Gribov:1962fw, Arbab:1969zr}. Factorization of residues enables one to determine the helicity structure at the photon vertex independently from the target, and conservation of parity reduces the number of helicity  components at each vertex. In the center-of-mass frame, the net helicity transfer between the vector meson and photon $|\lambda_\gamma - \lambda_V|$ can be $0$, $1$ or $2$, which we refer to as helicity conserving, single and double helicity flip respectively. Measurement of the photon spin-density matrix elements (SDMEs) can be used to determine the relative strength of these components. 

Spin-density matrix elements can be reconstructed from the angular distributions of the vector meson decay products~\cite{Schilling:1969um}. The first measurements of neutral vector meson SDMEs were performed at SLAC~\cite{Ballam:1972eq}, resulting in the following qualitative conclusions: the natural exchanges contributing to $\rho^0$, $\omega$ and $\phi$ production are predominantly helicity conserving, and the unnatural-parity contributions are negligible for $\rho^0$ production and consistent with a one-pion exchange for $\omega$ production. In this paper, we discuss the SLAC data in the context of a Regge-pole exchange model, which allows us to assess contributions of individual exchanges to the SDMEs. 
Various models have been proposed in the past~\cite{Drechsler:1966hhw, Schilling:1968awz, Daboul:1968oqf, Gotsman:1970dy, Ader:1970sg, Barker:1973dm, Laget:1994ba, Laget:2000gj, Sibirtsev:2003qh,Yu:2017vvp}, with different descriptions of the momentum-transfer dependence of the helicity amplitudes. In general these models reproduce  
 the differential cross sections, but lack  a detailed discussion of the implication of the Regge pole model for the SDMEs.

The paper is organized as follows.  In Section~\ref{sec:param}, we define the Regge amplitudes and discuss model parameters. In Section~\ref{sec:fit}, we discuss the fitting procedure. Specifically, we first 
 isolate the unnatural exchanges in  $\rho^0$ and $\omega$ production. We find that, within uncertainties, these components are consistent with  $\pi$ and $\eta$ exchanges so we neglect sub-leading trajectories. 
We  determine the residues of the dominant, natural exchanges by 
 the  $\gamma p$ and $\gamma d$ total cross sections. Using the SLAC data, the single and double helicity flip couplings are fitted to the three natural components of the SDMEs at the laboratory frame (target rest frame) photon beam energy of $E_\gamma = 9.3 \mbox{ GeV}$. The model is extrapolated to $E_\gamma = 2.8 \mbox{ GeV}$ and $4.7$ GeV and compared  to the three natural components of the SLAC SDMEs at these energies. In Section~\ref{sec:comp}, we compare the model to the nine $\omega$ and $\rho^0$ SDMEs obtained with a polarized beam at SLAC with $E_\gamma = 9.3 \mbox{ GeV}$, to the nine $\phi$ SDMEs from LEPS~\cite{Chang:2010dg} and Omega-Photon~\cite{Atkinson:1984cs}, and to the three $\omega$ SDMEs obtained with a unpolarized beam from CLAS~\cite{Williams:2009ab,Battaglieri:2002pr}, LAMP2~\cite{Barber:1985fr} and Cornell~\cite{Abramson:1976ks}.
 Furthermore, we test the Pomeron normalization for the $\omega$ and $\rho^0$ differential cross sections at $E_\gamma > 50$ GeV, and the Regge exchange normalization for the  $\omega$, $\rho^0$ and $\phi$ differential cross sections at $E_\gamma = 9.3 \mbox{ GeV}$ from Ref.~\cite{Ballam:1972eq}.  Lastly, we provide the predictions for the upcoming $\omega$, $\rho^0$ and $\phi$ SDMEs measurements in JLab experiments. In Section~\ref{sec:concl}, we summarize our findings and give conclusions.  Details regarding the relations between the frames (helicity, Gottfried-Jackson, $s$- and $t$-channel frames) are summarized in Appendix~\ref{app:frame}, the definition of the SDMEs are detailed in Appendix~\ref{app:sdme}, and further details on the amplitude parametrization are given in Appendix~\ref{app:HEL}.

\section{Regge model for vector meson photoproduction} \label{sec:param}
\begin{figure}[tb] 
\begin{center}
\includegraphics[width=\linewidth]{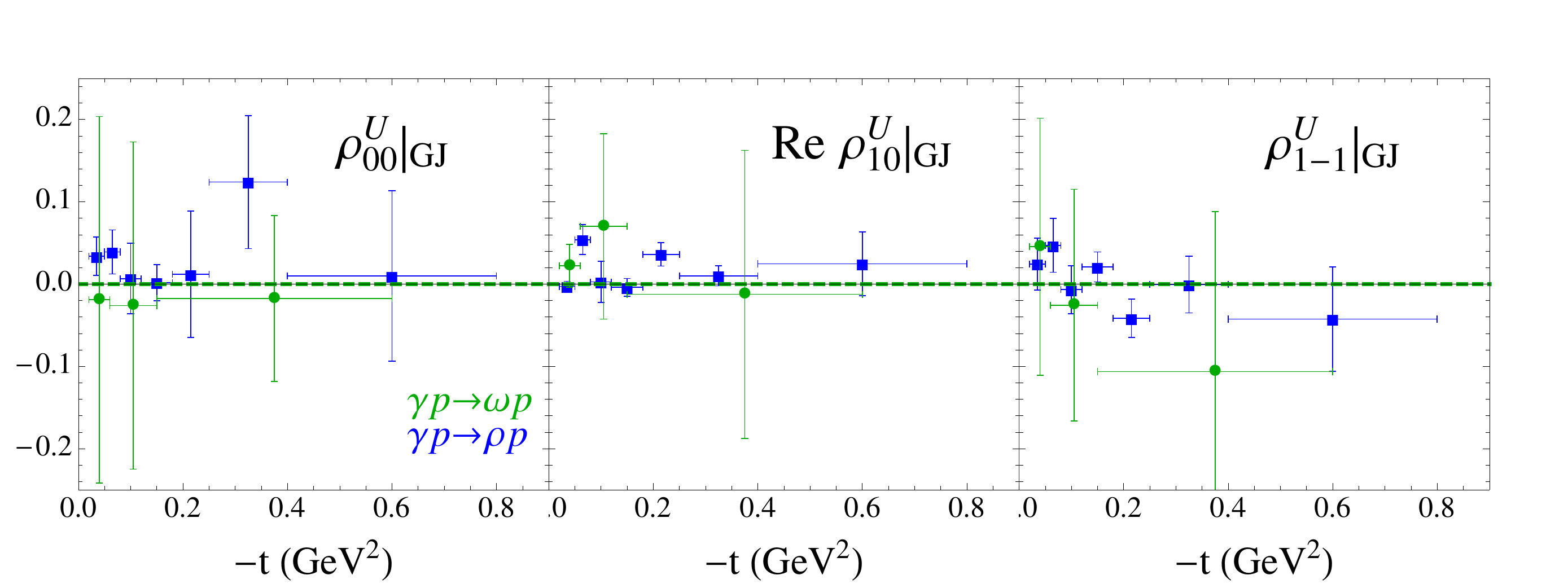}
\end{center}
\caption{\label{fig:sdme-unnat} Unnatural components of $\omega$ and $\rho^0$ SDMEs at $E_\gamma = 9.3$ GeV. The dashed lines are the theoretical expectation for a pseudoscalar exchange, $\rho^U_{00} = \rho^U_{10} = \rho^U_{1-1} = 0$. Data are taken from Ref.~\cite{Ballam:1972eq}.}
\end{figure}

At high energies, vector meson photoproduction is dominated by Pomeron and Regge exchanges. 
Regge exchanges can be characterized by the quantum numbers of the lowest spin meson on the trajectory, namely  
  isospin $I$, naturality $\eta = P(-1)^J$ (with the parity $P$), signature $\tau = (-1)^J$, charge conjugation $C$ and $G$-parity $G = C(-1)^I$. The leading trajectories contributing to vector meson photoproduction are 
\begin{align}\label{exchanges}  \nonumber
 &\ \ I^{G\eta \tau}J^{PC} & & \ \ I^{G\eta\tau}J^{PC} & & \ \ I^{G\eta\tau}J^{PC}\\  \nonumber
a_2 &: 1^{-++}2^{++} & \pi &: 1^{--+}0^{-+} & a_1&: 1^{---}1^{++}\\
f_2 &: 0^{+++}2^{++} & \eta &: 0^{+-+}0^{-+} & f_1&: 0^{+--}1^{++}
\end{align}
In addition to the exchanges in Eq.~\eqref{exchanges}, we also consider the natural-parity Pomeron exchange, which dominates at high energies. 
In the $\omega$ photoproduction model from~\cite{Yu:2017vvp},
a scalar exchange representing a $\sigma$ meson trajectory was also considered. Since the $\sigma$ meson trajectory is below the (leading) $f_2$ trajectory, we do not include it here.
Among all unnatural exchanges, the $\pi$ and $\eta$ trajectories are expected to dominate, since they are the closest to the scattering region. One can verify this by examining the SDMEs $\rho_{\lambda'\lambda}^U$, which in the Gottfried-Jackson (GJ) frame are determined by the unnatural exchanges (see Appendix \ref{app:frame}). The 
GJ frame is equivalent to the $t$-channel helicity frame where parity conservation implies a relation between  helicity amplitudes and the naturality of the exchanges.
Inspecting the SLAC data \cite{Ballam:1972eq},  one finds that the matrix elements $\rho^{U}_{00}|_{GJ}$, $\re \rho^{U}_{10}|_{GJ}$ and $\rho^{U}_{1-1}|_{GJ}$ for both $\omega$ and $\rho$ production are all consistent with zero. Moreover, the unnatural component of the differential cross section is compatible with a $\pi$-exchange model~\cite{Ballam:1972eq}. Hence, we assume that the unnatural components of the SDMEs are dominated by either $\pi$ or $\eta$ exchange. The $\eta$ exchange is introduced to describe the SDMEs in production of the $\phi$ meson, while its  contribution is negligible in $\omega$ and $\rho^0$ production. As we will see in Sec.~\ref{sec:param}, the normalization of these exchanges can be determined by vector meson radiative decays. Regarding axial vector exchanges, since the decay widths of $f_1, a_1 \to \gamma V$ are not known,\footnote{The only exception is for $\Gamma(f_1\to \gamma \phi) \sim 18$ keV.} their 
 contribution is difficult to evaluate. Within a specific quark model ~\cite{Yu:2017vvp}, the contribution of the $f_1$ to $\omega$ photoproduction is found to be negligible. As we will show, it is possible to saturate the unnatural components of the SDMEs by pseudoscalar exchanges. We therefore neglect the axial  vector trajectories. 
In summary, we consider the $s$-channel amplitudes in the form 
\begin{align}\label{eq:complete}
{\cal M}_{\substack{\lambda_V, \lambda_\gamma \\ \lambda', \lambda}}(s,t) & = \sum_{E}
{\cal M}^E_{\substack{\lambda_V, \lambda_\gamma \\ \lambda', \lambda}}(s,t), 
\end{align}
where the sum extends over the following $t$-channel reggeons: $E= \pi, \eta, {\mathbb P}, f_2, a_2$. 
  From the $s$-channel helicity amplitudes in Eq.~\eqref{eq:complete}, one can compute the SDMEs in the helicity or GJ frame  using Eqs.~\eqref{eq:rot}, and~\eqref{eq:SDMEs}, respectively. Assuming a factorized form for each exchange, 
\begin{align} \label{eq:factor}
{\cal M}^E_{\substack{\lambda_V, \lambda_\gamma \\ \lambda', \lambda}}(s,t) & = 
T^{E}_{\lambda_V \lambda_\gamma}(t)  R^E(s,t)  B^{E}_{\lambda' \lambda}(t),
\end{align}
where the top and bottom vertices  $T^E$ and $B^E$  describe the helicity transfer from the photon to the vector meson and between the nucleon target and recoil, respectively. 
According to Regge theory~\cite{Collins:1977jy}, the energy dependence factorizes into a power-law dependence $s^{\alpha_E(t)}$. The phase of the amplitude is determined by the signature factor $1+e^{-i\pi\alpha_E(t)}$, which is contained in $R^E$, 
\begin{subequations}
\begin{align}\label{eq:ReggePi}
R^U(s,t) & =  \frac{1+e^{-i\pi\alpha_U(t)}}{\sin \pi\alpha_U(t)}\hat s^{\alpha_U(t)} & U & = \pi, \eta \\
R^N(s,t) & =  \frac{\alpha_N(t)}{\alpha_N(0)}\frac{1+e^{-i\pi\alpha_N(t)}}{\sin \pi\alpha_N(t)}\hat s^{\alpha_N(t)} & N & = \mathbb P, f_2, a_2 .
\end{align}
\end{subequations}
We defined $\hat s = s/s_0$ with the scale chosen as $s_0 = 1$~GeV$^2$. We use a linear trajectory $\alpha_E(t) = \alpha_{E}(0) + \alpha_E' t$ for 
all exchanges.  The signature factor eliminates contributions from spin-odd poles induced by the denominator $\sin\pi\alpha_E(t)$. The factor $\alpha_E(t)/\alpha_E(0)$ simply removes the unphysical pole at $\alpha_E(t)=0$ that arises in the scattering region for the $f_2$ and $a_2$ exchanges. For consistency, we also include this factor for the Pomeron exchange, although the point $\alpha_{\mathbb P}(t)=0$ is far from the region of interest $-t\leq 1$ GeV$^2$. For the pseudoscalar exchanges, the pole at  $\alpha_{\pi,\eta}(t)=0$ is physical.

\subsection{Unnatural exchanges} \label{sec:unnat}
For unnatural exchanges $U=\pi,\eta$, the helicity structure of the photon vertex $T^U$ and the nucleon vertex $B^{U}$ 
 can be obtained by comparison with the high-energy limit of a single-particle exchange model. We obtain  (see Appendix~\ref{app:HEL}), 
\begin{subequations}\label{eq:pion}
\begin{align} \nonumber
T^U_{\lambda_V \lambda_\gamma}(t) & =  \\  \beta^U_{\gamma V} 
 \bigg( & \lambda_\gamma\delta_{\lambda_V, \lambda_\gamma} 
- \sqrt{2} \frac{ \sqrt{-t} }{m_V}\delta_{\lambda_V,0}+ 
 \frac{-t}{m_V^2} \lambda_\gamma \delta_{\lambda_V,-\lambda_\gamma} \bigg),\\
B^U_{\lambda' \lambda}(t) &= \beta^{U}_{pp}\left(  \delta_{\lambda,-\lambda'} \frac{\sqrt{-t}}{2m_p} \right),
\end{align}
\end{subequations}

with $m_V$ and $m_p$ being the vector meson and nucleon masses, respectively. 
The residues $\beta^U_{\gamma V}$ and $\beta^{U}_{pp}$ are determined from the radiative decay widths $\Gamma(V\to \gamma\pi)$, $\Gamma(V\to \gamma\eta)$ and the nucleon couplings $g_{\pi pp}$, $g_{\eta pp}$, respectively. The overall nonflip couplings of the reaction are written  $\beta^U_{0,V} = \beta^{U}_{\gamma V} \beta^{U}_{pp}$.\footnote{The index $0$ stands for the helicity difference at the top vertex, $|\lambda_\gamma-\lambda_V|=0$.} The details of the calculation are given in Appendix~\ref{app:HEL}. The unnatural trajectory is $\alpha_U(t) = \alpha'_U(t-m_\pi^2)$ with $\alpha_U' = 0.7$ GeV$^{-2}$.
The parameters for the unnatural exchanges are summarized in Table~\ref{tab:unnat}.
The photon vertex $T^{U}_{\lambda_V \lambda_\gamma}$ involves all possible helicity structures, with each unit of helicity flip contributing a factor of $\sqrt{-t}$.
Because of charge conjugation, there is only one helicity structure at the the nucleon vertex, the helicity flip, which corresponds to the factor $\delta_{\lambda,-\lambda'} \sqrt{-t}/2m_p$. 

\begin{table}[htb]
\centering
\caption{Model parameters for the unatural exchanges. The parameters $\alpha'_U$ are expressed in GeV$^{-2}$.
\label{tab:unnat}}
\begin{tabular}{c|cc }
$U$  & $\pi$ & $\eta$ \\
\hline \rule{0pt}{2.6ex}
$\beta^U_{0,\omega}$ & 3.11 & 0.36  \\
$\beta^U_{0,\rho}$ & 1.11 & 0.10  \\
$\beta^U_{0,\phi}$ &  0.30 & 0.27 \\
$\alpha_U(0)$ &  $-$0.013 & $-$0.013 \\
$\alpha'_U$ & 0.7   & 0.7  \\
\end{tabular}
\end{table}

\subsection{Natural exchanges} \label{sec:nat}
The trajectories of the natural exchanges are known and we use~\cite{Irving:1977ea,Collins:1977jy}
\begin{subequations}
\begin{align}
\alpha_{\mathbb P}(t) &= 1.08 + 0.2 \,t/\text{GeV}^{2}, \\
\alpha_{f_2 , a_2}(t) &= 0.5 + 0.9 \,t/\text{GeV}^{2}.
\end{align}
\end{subequations}
For natural exchanges, $N = \mathbb P,f_2, a_2$. The top vertex involves three  helicity components: a helicity nonflip, single flip and double flip. As for unnatural exchanges, each of these comes with an appropriate power of the factor $\sqrt{-t}/m_V$,
\begin{align} \label{eq:topVnat} \nonumber
T^{N}_{\lambda_V \lambda_\gamma}(t) & =  \beta^{N}_{\gamma V} e^{b_N t} \\
 \times \bigg( &\delta_{\lambda_V,\lambda_\gamma} + \beta^N_1 \frac{\sqrt{-t}}{m_V}  \frac{\lambda_\gamma}{\sqrt{2}} \delta_{\lambda_V,0} +\beta^N_2
 \frac{-t}{m^2_V}  \delta_{\lambda_V,-\lambda_\gamma}  \bigg).
\end{align}
To be consistent with factorization, and to reduce the number of parameters, we assume that the couplings $\beta_1^N$ and $\beta_2^N$ are the same for all vector mesons.  
The steep falloff of the forward differential cross section is well described by exponential factors, gamma functions~\cite{Irving:1977ea,Laget:2000gj} or dipole form factors~\cite{Donnachie:1983hf,Laget:1994ba, Oh:2000zi,Sibirtsev:2003qh,Yu:2017vvp}. All of these models can be approximated by an exponential function of the form $e^{b_N t}$~\cite{Schilling:1968awz, Gotsman:1970dy, Ader:1970sg, Barker:1973dm}.  We obtain $b_{\mathbb P}=3.6$ GeV$^{-2}$ by approximating the form factors from~\cite{Donnachie:1983hf}, and $b_{a_2}=0.53$ GeV$^{-2}$ and $b_{f_2}=0.55$ GeV$^{-2}$ by approximating the $t$-dependence of the $a_2$ and $f_2$ poles with a Breit-Wigner line shape as described in Appendix~\ref{app:HEL}. 
For the nucleon vertex we include the two possible helicity combinations, a nonflip and single flip, 
\begin{align} \label{eq:bottomVnat}
B^{N}_{\lambda' \lambda}(t) & = \beta^{N}_{pp} \left( \delta_{\lambda,\lambda'}  + 2 \lambda \kappa_N \frac{\sqrt{-t}}{2m_p} \delta_{\lambda,-\lambda'} \right).
\end{align}
The SDMEs probe the helicity structure of the photon vertex. They are weakly dependent on the helicities  
 at the  nucleon vertex. On the contrary, The helicity flip couplings $\kappa_N$ thus play a minor role in our analysis. Moreover isoscalar exchanges, {\it e.g.}, the $f_2$ and Pomeron, are empirically helicity nonflip at the nucleon vertex~\cite{Irving:1977ea}. Therefore, we set $\kappa_{f_2}=\kappa_{\mathbb P}=0$. The isovector exchanges are empirically helicity flip dominant. We model this feature by  using $\kappa_{a_2} = 8.0$~\cite{Irving:1977ea}.

The special nature of the Pomeron prevents us from computing its overall normalization $\beta_{0,V}^{\mathbb P} = \beta^{\mathbb P}_{\gamma V} \beta^{\mathbb P}_{pp}$ by using radiative decays. 
We thus determine the normalization $\beta_{0,V}^N = \beta^{N}_{\gamma V} \beta^{N}_{pp}$ by fitting the $\gamma p$ and $\gamma d$ total cross sections and invoking vector meson dominance (VMD). We first relate the overall normalizations $\beta_{0,V}^N$ to the $\gamma p$ and $\gamma d$ total cross section. Using the optical theorem, our Regge parametrization in~\eqref{eq:factor} leads to
\begin{align} \nonumber
\sigma(\gamma p ) & = \frac{1}{2m_pE_\gamma} \bigg(\beta^{\gamma \gamma}_{\mathbb P} \hat s^{\alpha_{\mathbb P}(0)} + \beta^{\gamma \gamma}_{f_2}\hat s^{\alpha_{f_2}(0)}  + \beta^{\gamma \gamma}_{a_2}\hat s^{\alpha_{a_2}(0)} \bigg), \\
\sigma(\gamma d) & = \frac{1}{2m_pE_\gamma}\left(2\beta^{\gamma \gamma}_{\mathbb P} \hat s^{\alpha_{\mathbb P}(0)} +2 \beta^{\gamma \gamma}_{f_2} \hat s^{\alpha_{f_2}(0)} \right).
\label{eq:sigtot}
\end{align} 

The factors  $\beta^{\gamma \gamma}_{N} $ represent couplings of the natural exchange $N$ in the forward scattering direction $\gamma p \to \gamma p$.  We need to relate these factors, via VMD, to the factors $\beta_{0,V}^N$ appearing in vector meson photoproduction. In order to use VMD, we use the following interaction between photon field $A_\mu$ and the vector meson fields~\cite{GellMann:1961tg,GellMann:1962jt,GellMann:1962xb}:
\begin{align}\label{eq:VMD}
{\cal L} & = - eA^\mu \left( \frac{m^2_\rho}{\gamma_\rho} \rho_\mu 
+  \frac{m^2_\omega}{\gamma_\omega} \omega_\mu 
+ \frac{m^2_\phi}{\gamma_\phi} \phi_\mu \right).
\end{align}
From this interaction,\footnote{The $\gamma_V$ couplings can be cast in terms of the vector meson decay constants $\langle 0| \sum_{q =u,d,s} e_q \bar q \gamma_\mu q(0) | V(\epsilon,P)\rangle = f_V \epsilon^V_\mu(P)= (m_V^2/\gamma_V) \epsilon^V_\mu(P) $.} and neglecting the electron mass, one finds for the electronic decay width $\Gamma(V \to e^+ e^-) = m_V (\alpha^2/3) (4\pi/\gamma^2_V)$, which determines the couplings $\gamma_V$ that we tabulate in Table~\ref{tab:vmd}. 
\begin{table}[htb]
\centering
\caption{Vector meson dominance parameters.
\label{tab:vmd}}
\begin{tabular}{c|cc}
$V$  &$\Gamma(V\to e^+e^-)$ & $4\pi/\gamma_V^2$  \\
\hline
$\rho^0$ & 7.04(6) keV & 0.506(4)  \\
$\omega$ & 0.60(2) keV & 0.044(1)  \\
$\phi$ &  1.26(1) keV& 0.070(1)  
\end{tabular}
\end{table} 
The SU(3) quark model predictions $\gamma_\omega/\gamma_\rho=3$ and $\gamma_\omega/\gamma_\phi=-\sqrt{2}$ compare well with the VMD predictions, $\gamma_\omega/\gamma_\rho=3.4(6)$ and $\gamma_\omega/\gamma_\phi=-1.3(1)$. However, it is well known that the $\phi$ meson differential cross section produces a value of $\gamma_\phi$ that is twice as large as the one obtained from the leptonic decay width~\cite{Ballam:1972eq}. For consistency, we will use the $\gamma_\phi$ value obtained from the leptonic decay width, but we keep an eye on this  discrepancy when comparing to the data.

Assuming that the Pomeron has a gluonic nature and therefore has couplings which are independent of the quark flavor~\cite{Donnachie:1994zb}, we derive the relation between the total cross section couplings in Eq.~\eqref{eq:sigtot} and the overall normalization of the Pomeron $\beta^{\mathbb P}_{0,V}$ in our model for vector meson photoproduction,
\begin{subequations} \label{eq:gV}
\begin{align}\label{eq:gPom}
\beta^{\mathbb P}_{0,V} & = \beta_{\mathbb P}^{\gamma\gamma} \frac{e}{\gamma_V} \times \left( \frac{e^2}{\gamma_\rho^2} + \frac{e^2}{\gamma_\omega^2} + \frac{e^2}{\gamma_\phi^2}\right)^{-1}.
\end{align}
We note that by increasing $\gamma_\phi$ by a factor of two, the $\omega$ and $\rho^0$ couplings of the Pomeron would change by only 10$\%$. 
For the Regge exchanges, we assume ideal mixing for vector and tensors mesons and extract the remaining couplings using vector meson dominance:
\begin{align}
\beta^{f_2}_{0,\omega/\rho} & =  \beta_{f_2}^{\gamma\gamma} \frac{e }{\gamma_{\omega/\rho}} \times \left(\frac{e^2}{\gamma_\rho^2} + \frac{e^2}{\gamma_\omega^2}\right)^{-1},  \\
\beta^{a_2}_{0,\omega/\rho} & =  \beta_{f_2}^{\gamma\gamma}\frac{\gamma_{\omega/\rho}} 
{2 e}, \\
\beta_{\gamma \phi}^{f_2}&=\beta_{\gamma \phi}^{a_2}=0.
\end{align}
\end{subequations}

\begin{figure}[htb]
\begin{center}
\includegraphics[width=\linewidth]{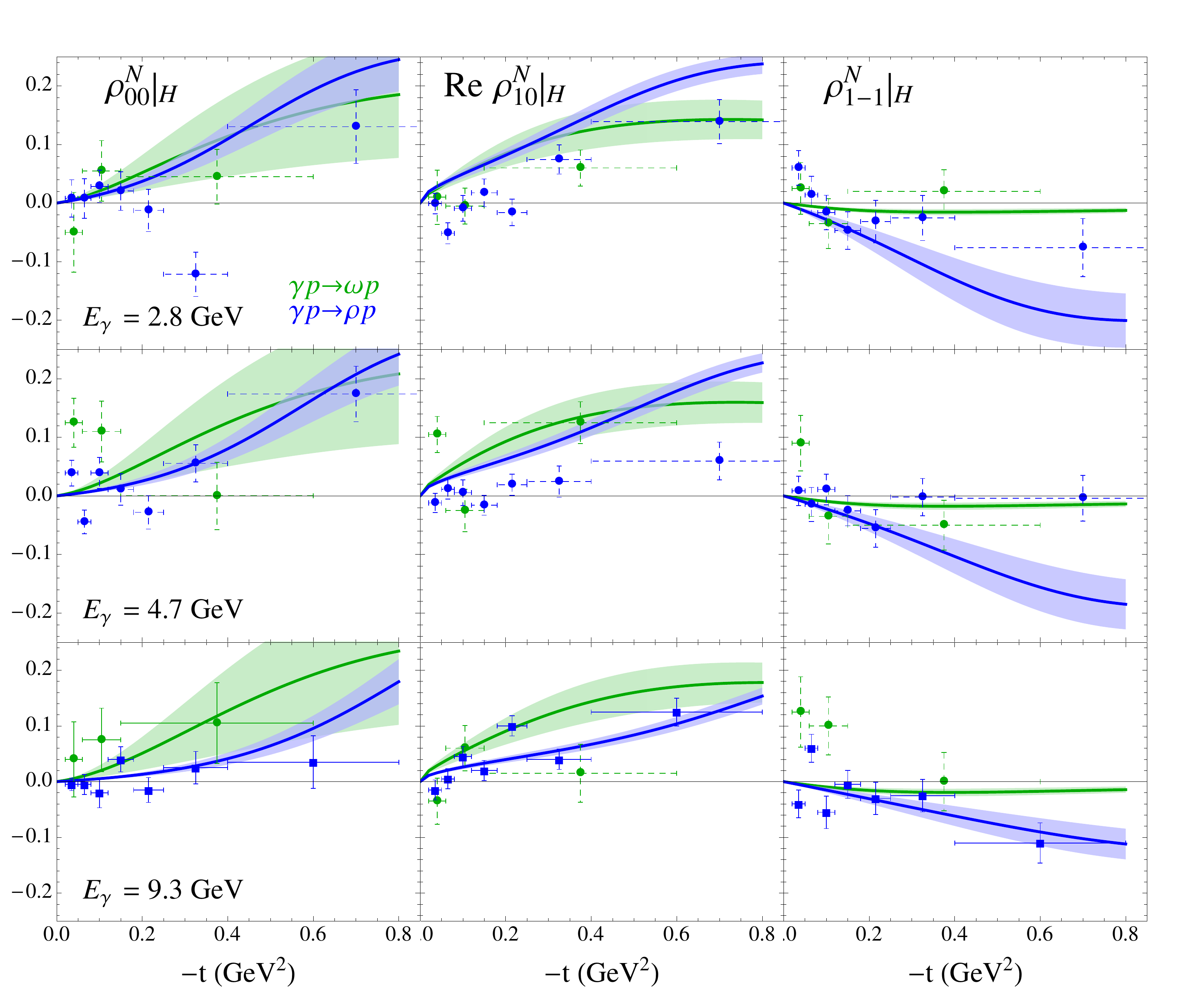}
\end{center}
\caption{\label{fig:sdme-nat-slac}Natural components of $\omega$ and $\rho^0$ photoproduction SDMEs at $E_\gamma = 2.8, 4.7$ and $9.3$ GeV. The lines are our model, determined by the 9.3 GeV data only and extrapolated to lower energies. The dashed points are not included in the fitting procedure. 
The data are taken from Ref.~\cite{Ballam:1972eq}. }
\end{figure}

We choose to determine the helicity couplings $\beta_1^N$ and $\beta_2^N$ through a fit to the SLAC data. 
Since our formalism is based on a high-energy expansion, we determine the parameters only with the highest energy bin. Specifically, we inspect the natural components of the SDMEs at $E_\gamma = 9.3$ GeV. Assuming only one natural exchange $N$, our form in Eq.~\eqref{eq:topVnat} for the top vertex leads to
\begin{subequations} \label{eq:Nsdme}
\begin{align}
\rho^N_{00}(s,t) & = \frac{\left(\beta_1^N\right)^2}{A(t)}  \frac{-t}{m_V^2}, \\
\re \rho^N_{10}(s,t) & = \frac{\beta^N_1}{2A(t)}\frac{\sqrt{-t}}{m_V} \left(1+\beta^N_2  \frac{-t}{m_V^2}\right),  \\
\rho^N_{1-1}(s,t) & =\frac{\left(\beta_2^N\right)^2}{A(t)}  \frac{-t}{m_V^2}, 
\end{align}
\end{subequations}
with $A(t) = 1-\left(\beta_1^N\right)^2 t/m_V^2 + \left(\beta_2^N\right)^2 t^2/m_V^4$. The factorization hypothesis in Eq.~\eqref{eq:factor} and the conservation of angular momentum implies the vanishing of these SDMEs in the forward direction. This is indeed observed in all of the $\rho^0$ SDMEs, but is inconsistent with the $\rho^N_{1-1}$ elements for $\omega$ photoproduction as seen in Fig.~\ref{fig:sdme-nat-slac}. The expressions in Eq.~\eqref{eq:Nsdme} also tell us that we should expect $|\rho^N_{00}|  < |\re \rho^N_{10}|$ for small $t$. Again, this relation is satisfied for $\rho^0$ photoproduction but seems to be violated for $\omega$ photoproduction. The element $\rho^N_{00}$ is significantly larger for $\omega$ photoproduction compared to $\rho^0$ photoproduction, suggesting a larger single-helicity flip for the isovector exchange. The deviation from zero observed in the elements $\re \rho^N_{10} $ and $\rho^N_{1-1} $ for $\rho^0$ photoproduction suggests a nonzero single and double helicity flip for the isoscalar exchanges. We associate these couplings with the $f_2$ exchange and keep the Pomeron helicity conserving as is often assumed. This hypothesis could be checked with $\phi$ photoproduction as we will discuss later. According to our discussion we impose $\beta_1^{\mathbb P} = \beta_2^{\mathbb P}= \beta_2^{a_2}=0$ and thus need to fit the helicity couplings $\beta_1^{f_1}, \beta_1^{f_2}, \beta_1^{a_2}$.

\section{Fitting procedure} \label{sec:fit}
\begin{table}[t]
\centering
\caption{Model parameters for the natural exchanges. The parameters $b_N$ and $\alpha'_N$ are expressed in GeV$^{-2}$. The $\beta^N_{\{0,1,2\},V}$ parameters are calculated using the fit discussed in Section~\ref{sec:fit}; the other parameters are estimated or discussed in Section~\ref{sec:param}.
\label{tab:nat}}
\begin{tabular}{c|ccc }
$N$  &$\mathbb P$ & $f_2$ & $a_2$ \\
\hline \rule{0pt}{2.6ex}
$\beta^N_{0,\omega}$ & 0.739(1) & 0.730(10)  & 1.256(85) \\
$\beta^N_{0,\rho}$ & 2.506(5) & 2.476(34)  & 0.370(25)\\
$\beta^N_{0,\phi}$ &  0.932(2)& 0  & 0 \\
$\beta^N_1$ & 0  & $\phantom{+}$0.95(19)  & 0.83(34)\\
$\beta^N_2$ & 0 &$-$0.56(17)  & 0\\ \hline
$\kappa_N $ & 0 & 0 & 8.0 \\
$b_N $ & 3.60 & 0.55 & 0.53 \\
$\alpha_N(0)$ &  1.08 & 0.5 & 0.5 \\
$\alpha'_N$ & 0.2   & 0.9 & 0.9  \\
\end{tabular}
\end{table}

We determine the six couplings $\beta^{\gamma\gamma}_{\mathbb P}$, $\beta^{\gamma\gamma}_{f_2}$, $\beta^{\gamma\gamma}_{a_2}$, $\beta_1^{f_1}$, $\beta_1^{f_2}$, $\beta_1^{a_2}$ using  a combined fit of the $\gamma p$ and $\gamma d$ total cross sections from the Review of Particle Physics~\cite{pdg} for $E_\gamma>2$ GeV,  the three $\rho^0$ natural exchange SDMEs ($\rho^N_{00}$, $\re \rho^N_{10}$ and $\rho^N_{1-1}$) and the element $\rho^N_{00}$ for $\omega$ photoproduction at $E_\gamma= 9.3$ GeV obtained at SLAC~\cite{Ballam:1972eq}. We do not include the two other natural components of the SDMEs in $\omega$ photoproduction as they are inconsistent with our working hypothesis. The fit of the total cross sections and the fit of the SLAC SDMEs are combined in a single fit. There are 308 (total cross sections) plus 24 (SDMEs) data points and six fit parameters. The other model parameters ($b_N$, $\kappa_N$, $\gamma_V$ and the $\pi$- and $\eta$-exchange couplings) are kept fixed at values discussed in the previous section.  The expressions for the natural components of the SDMEs used in the fit is given in Eqs~\eqref{eq:SDMEs} and \eqref{eq:sdmeNat-unnat}. The fit results in the reduced $\chi^2/\text{d.o.f.}$ of 1.96 (1.84 for the total cross sections and 0.12 for the SDMEs),  and the fitted parameters are
\begin{subequations}
\label{eq:fitcouplings}
\begin{align}
\beta_{\mathbb P}^{\gamma \gamma} & = 0.187(1) & \beta_1^{f_2} & =  0.95(19)\\
\beta_{f_2}^{\gamma \gamma} & = 0.164(2) & \beta_2^{f_2} & =  -0.56(17)\\
\beta_{f_2}^{\gamma \gamma} & = 0.045(3) &\beta_1^{a_2} & =  0.83(34) .
\end{align}
\end{subequations}
The photon couplings are extracted from Eqs~\eqref{eq:gV}. The parameters of the exchanges calculated from Eq.~\eqref{eq:fitcouplings} for vector meson photoproduction are summarized in Table~\ref{tab:nat}. 


\begin{figure}[htb]
\begin{center}
\includegraphics[width=0.85\linewidth]{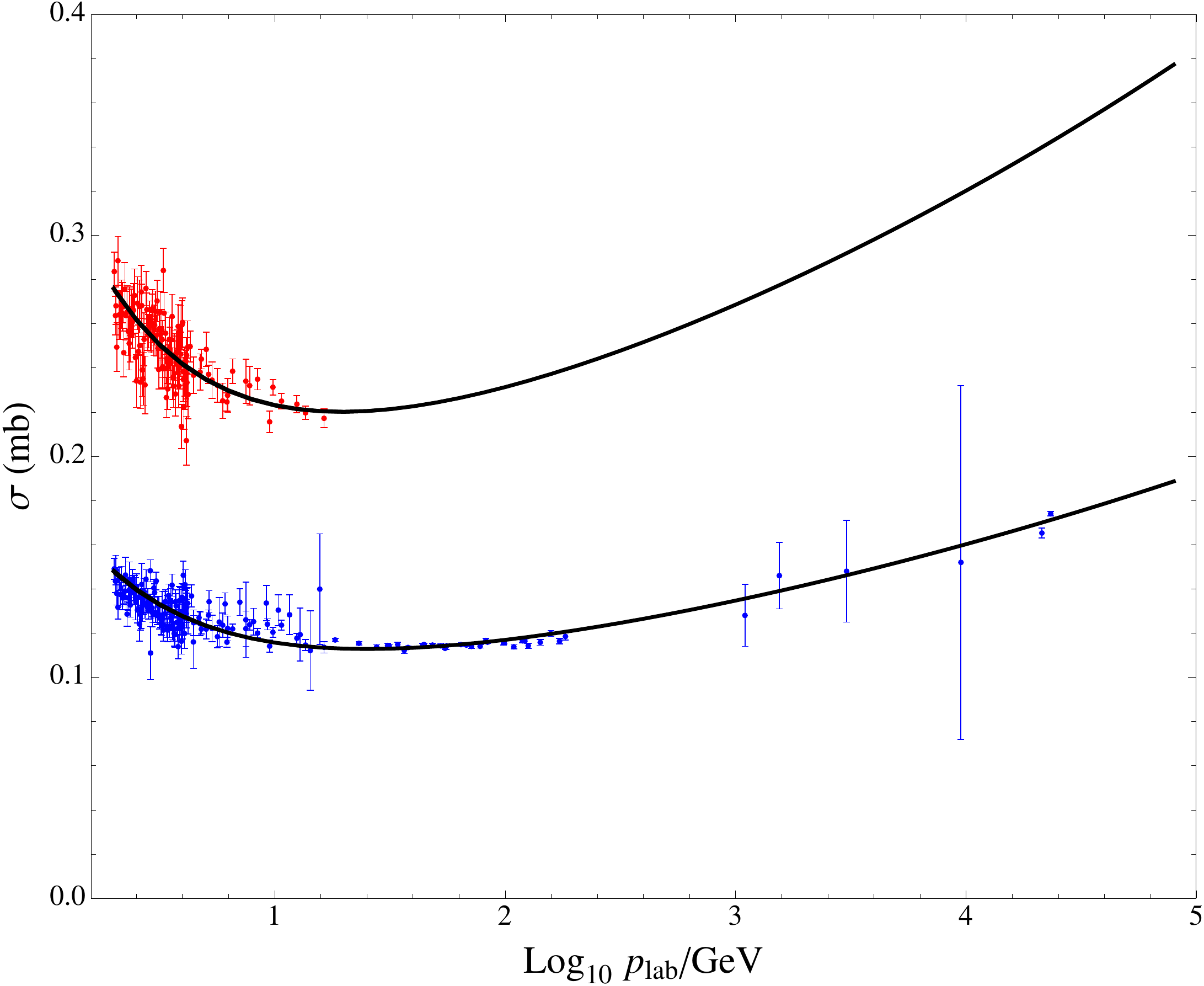}
\end{center}
\caption{\label{fig:totalcross}Total cross section $\sigma(\gamma p)$ (blue) and $\sigma(\gamma d)$ (red). The black lines are the results of our fit (the thickness of the lines represent the error band). The data are taken from Ref.~\cite{pdg}.}
\end{figure}

\section{Comparison with data} \label{sec:comp}
As we discussed above, the SDMEs for $\rho^0$ photoproduction are more consistent with our model for diffractive production than for $\omega$ photoproduction. This can be observed in Fig.~\ref{fig:sdme-nat-slac}. The bands on the figures represent one standard deviation from our model.  The wider band in the $\omega$ model originates from the stronger dominance of the Regge exchanges, whose normalizations are less constrained by the total cross sections. The Pomeron normalization is indeed more constrained and yields a smaller uncertainty in the $\rho^0$ model. We have also included the data at $E_\gamma= 4.7$ and $2.8$ GeV from SLAC in Fig.~\ref{fig:sdme-nat-slac}. They compare well to our model evaluated at these lower energies. 

\begin{figure}[htb]
\begin{center}
\includegraphics[width=\linewidth]{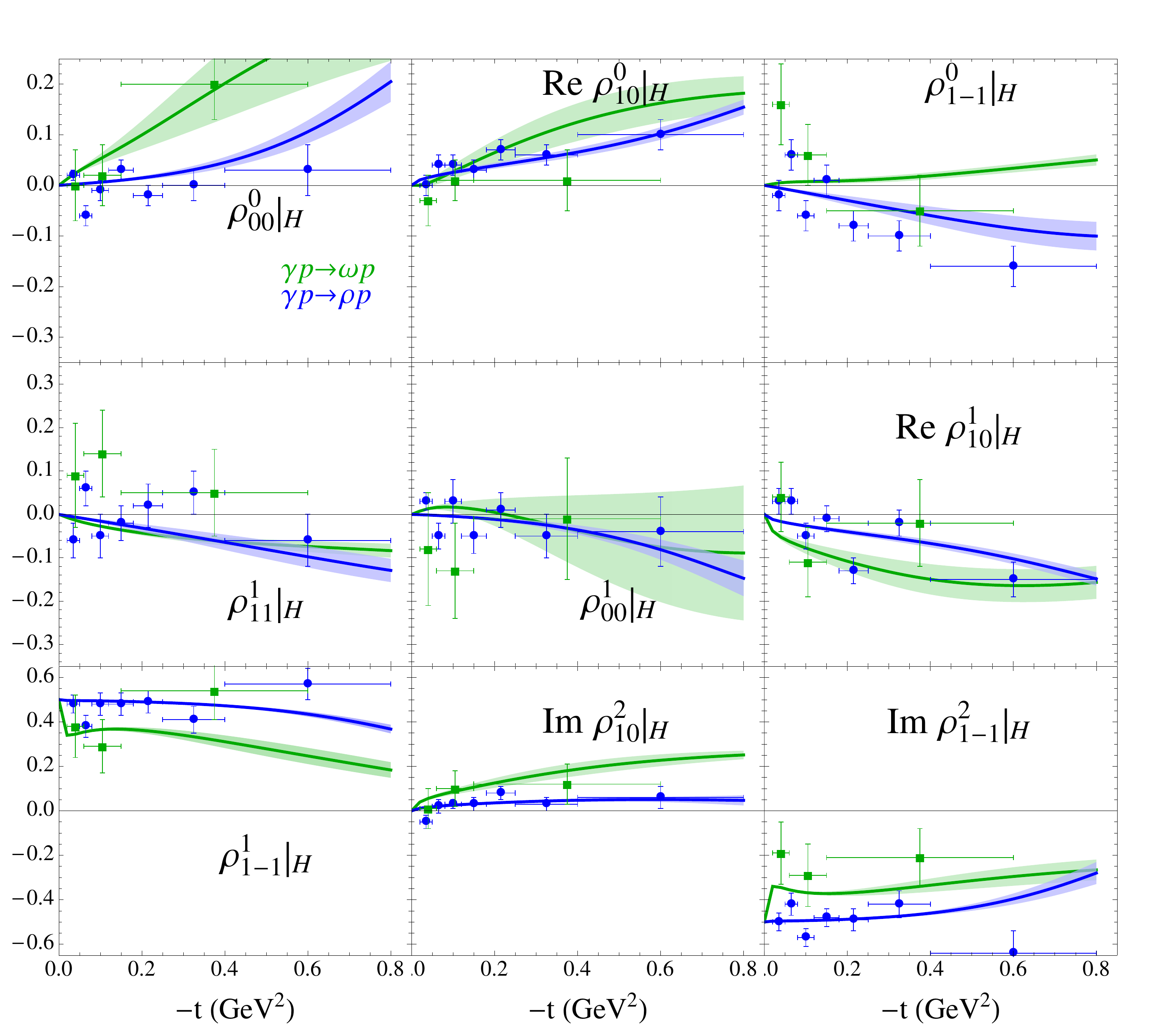}
\end{center}
\caption{\label{fig:sdme-all-slac93}Comparison between our model and $\omega$ and $\rho^0$ SDMEs at $E_\gamma = 9.3$ GeV.  The data are taken from Ref.~\cite{Ballam:1972eq}.}
\end{figure}

In Fig.~\ref{fig:sdme-all-slac93},  we present the comparison between the $\omega$ and $\rho^0$ models and the SLAC data at 9.3 GeV for all nine SDMEs. There is a general agreement between the model and the data, but we wish to discuss some inconsistencies. The elements in the bottom panels $\rho_{1-1}^1$, $\im \rho^2_{10}$ and $\im \rho^2_{1-1}$ were not included in the fitting but are nevertheless well described by the model. In particular, we note the dominance of the natural exchanges in $\rho_{1-1}^1$ and $\im \rho^2_{1-1}$ in the case of $\rho^0$ photoproduction with small deviation for the $\omega$ case, as expected from the stronger $\pi$ exchange. The main noticeable discrepancy arises in $\rho^1_{11}$ for $\omega$ photoproduction. Since the pseudoscalar exchanges are smaller than the natural exchanges, we would expect $\rho_{11}^1 \sim \rho_{1-1}^0$. The data does not display this feature and thus our model does not describe $\rho^1_{11}$ well. Furthermore, since the contribution from the $\pi$ exchange to $\rho^1_{11}$ is negative (see Appendix~\ref{app:HEL}), we would expect $\rho_{11}^1 < \rho_{1-1}^0$, which is featured in our $\omega$ model but not in the SLAC data. The sign of the element $\rho^1_{11}$ would be an important check for our model when GlueX data becomes available. 

\begin{figure}[htb]
\begin{center}
\includegraphics[width=\linewidth]{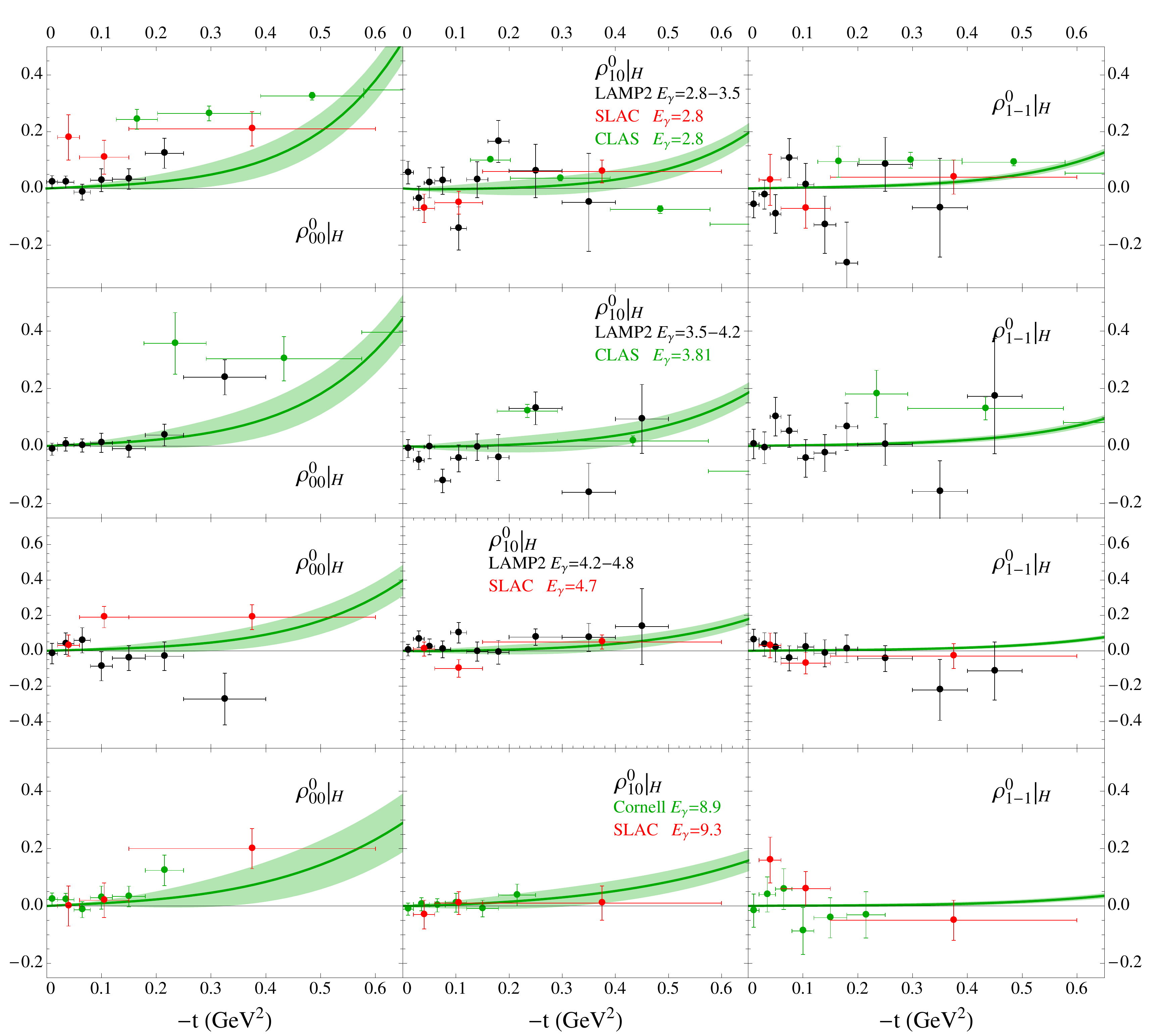}
\end{center}
\caption{\label{fig:sdme-all-ome}Unpolarized SDMEs for $\omega$ photoproduction. The lines are our model. $E_\gamma$ is the beam energy in the laboratory frame in GeV. The data are taken from SLAC~\cite{Ballam:1972eq}, CLAS~\cite{Williams:2009ab,Battaglieri:2002pr}, LAMP2~\cite{Barber:1985fr} and Cornell~\cite{Abramson:1976ks}.
}
\end{figure}

Although our model has been constrained at $E_\gamma = 9$ GeV, we present in Fig.~\ref{fig:sdme-all-ome} the comparison between our model and the unpolarized SDMEs at lower energies. The extrapolation to lower energies is in principle not in the range of applicability of the Regge-pole approximation. Despite the significant uncertainties in all the presented data sets, we conclude that our extrapolated model describes the lower energies data sets fairly well. It is also worth noting that the data from Ref.~\cite{Abramson:1976ks} at $E_\gamma=8.9$ GeV are consistent with our factorization hypothesis, {\it i.e.}, $\rho^0_{1-1} \sim -t$ in the forward direction. We conclude that the SLAC data may suffer from large errors. The forthcoming measurement by the GlueX collaboration could confirm the factorization of the vector meson production, {\it i.e.}, $\rho_{1-1}^0(t) \sim -t$ in the forward direction at high energies.

\begin{figure}[htb]
\begin{center}
\includegraphics[width=\linewidth]{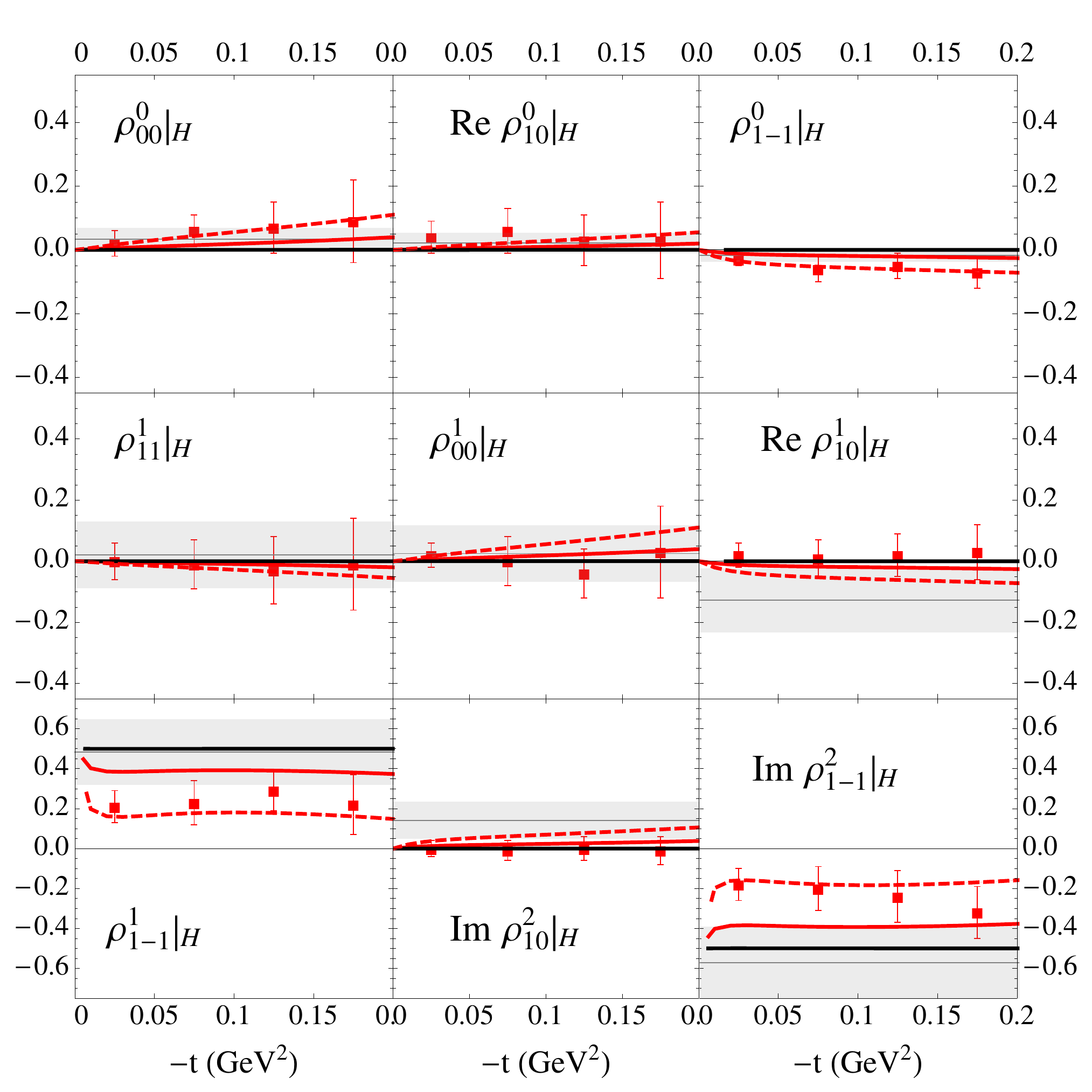}
\end{center}
\caption{\label{fig:sdme-phi-wu}The SDMEs of $\phi$ photoproduction at $E_\gamma = 2.17-2.37$ GeV (red squares) from Ref.~\cite{Chang:2010dg} and at $E_\gamma = 20-40$ GeV (gray band) from Ref.~\cite{Atkinson:1984cs} (SDMEs integrated over $t$ represented as a band over the $t$ range). The lines are our model at $E_\gamma = 2.27$ GeV (solid red with $\beta_{0,\phi}^{\mathbb P}= 0.932$ and dashed red with $\beta_{0,\phi}^{\mathbb P}= 1/2 \times 0.932$ for the Pomeron coupling) and $E_\gamma = 30$ GeV (black).}
\end{figure}

Our model simplifies for $\phi$ photoproduction. In this case we simply neglect the $f_2$ and $a_2$ Regge exchanges, as they are not expected to couple to $\gamma \phi$ if one assumes perfect mixing. The relevant exchange would then be the $f_2'$, the hidden strangeness  partner of the $f_2$. However, its intercept, and therefore its overall strength, is smaller due its higher mass. We neglect this contribution and assume that the only relevant natural contribution is provided by the Pomeron. Since our Pomeron is purely helicity conserving, the SDMEs are very simple at high energies. The only non-zero components are $\rho^1_{1-1} = - \im \rho^2_{1-1} = 1/2$. This picture is consistent with the SLAC measurement at $9.3$ GeV~\cite{Ballam:1972eq}. In Fig.~\ref{fig:sdme-phi-wu}, we compare our model to the data from the Omega-Photon collaboration~\cite{Atkinson:1984cs}. Their data are taken in the energy range $E_\gamma = 20 - 40$ GeV. They are consistent with the SLAC data but have somewhat smaller uncertainties. We also extrapolated our model to $E_\gamma = 2.27$ GeV to compare with the data from the LEPS collaboration~\cite{Chang:2010dg}. At lower energies, we observe deviations from pure helicity conservation, {\it i.e.}, deviation from $\rho^1_{1-1} = - \im \rho^2_{1-1} = 1/2$. This is triggered by unnatural exchanges. Since the $\pi$ couples weakly to $\gamma \phi$, we included $\eta$ exchange in our model. The very small coupling $g_{\phi\gamma\pi}$, inferred from radiative decays, cannot solely explain the deviation from helicity conservation in the elements $\rho^1_{1-1}$ and $\im \rho^2_{1-1}$ at $E_\gamma = 2.27$ GeV. The inclusion of $\eta$ exchange increases the relative importance of unnatural exchange. We should also note that we considered the $\eta$ degenerate with the $\pi$. With the $\eta$ pole being further from the scattering region, the factor $\alpha' \pi / \sin \pi \alpha_\eta(t) \sim 1/(m_\eta^2-t)$ is not strong enough to trigger the depletion close to the forward direction in $\rho^1_{1-1}$ and  $\im \rho^2_{1-1}$. Nevertheless, the SDMEs from the LEPS collaboration indicate an even larger relative strength of unnatural vs. natural exchanges than in our model. As we pointed out, the Pomeron coupling $g_{\mathbb P }^{\gamma \phi}$ from the $\phi$ meson leptonic width and VMD is overestimated. The relative strength of the unnatural exchanges in the SDMEs are thus underestimated. We illustrate the effect of reducing the Pomeron coupling by a factor of two in Fig.~\ref{fig:sdme-phi-wu}. The dashed red line, obtained with $\beta_{0,\phi}^{\mathbb P} = 1/2 \times 0.932$, leads to a better agreement with the data. Alternatively, we could have increased the coupling $g_{\eta NN}$. As we discussed in Ref.~\cite{Nys:2016vjz}, the $\eta$ coupling to the nucleon is not known very precisely.  From the investigation of $\phi$ SDMEs at $E_\gamma=2.27$ GeV, we conclude that the ratio of natural and unnatural component is $\beta_{0,\phi}^{N}/\beta_{0,\phi}^{U} =  0.266$.

\begin{figure}[htb]
\begin{center}
\includegraphics[width=\linewidth]{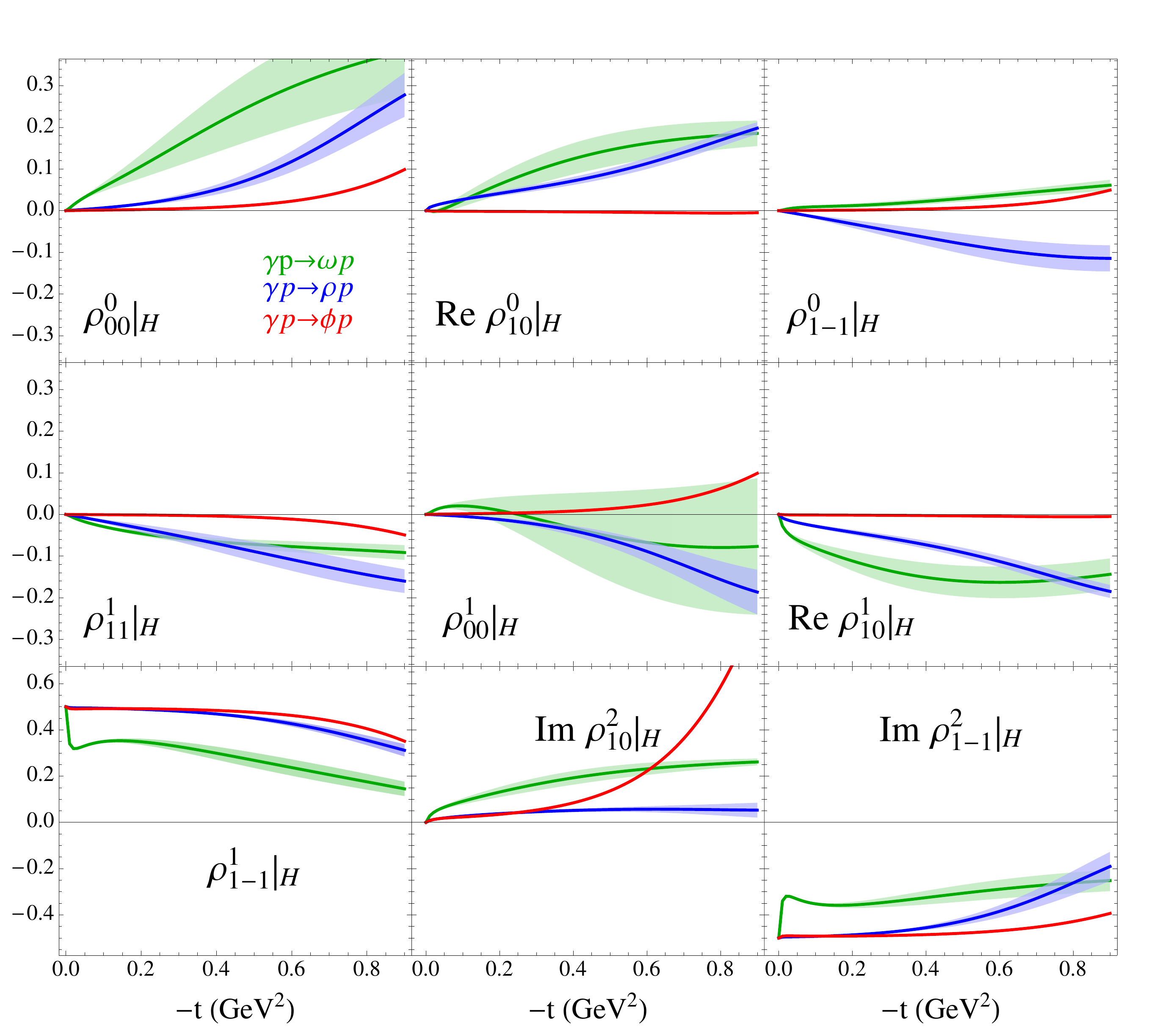}
\end{center}
\caption{\label{fig:sdme-all-gluex}The SDMEs of $\omega$, $\rho^0$ and $\phi$ photoproduction at $E_\gamma = 8.5$ GeV, the average polarized beam energy in the laboratory frame. }
\end{figure}

Our prediction for $\omega, \rho^0$ and $\phi$ vector meson photoproduction at GlueX is displayed in Fig.~\ref{fig:sdme-all-gluex}. We used $E_\gamma = 8.5$ GeV, the average beam energy with polarization. As already commented, the  bulk of the uncertainties in our model come from Regge exchanges. It is therefore not surprising that the uncertainties in the $\phi$ meson SDMEs are very small. The bending of the curves as $|t|$ increases in our $\phi$ model originate from the pseudoscalar exchanges. We have not included an exponential falloff in their parametrization. Therefore, their effects can be observed away from the forward direction where the natural exchanges are exponentially suppressed. If the $\phi$ SDMEs remain flat in a larger $t$ range, one would just need to incorporate an exponential falloff in the $\eta$ exchange. 

\begin{figure}[htb]
\begin{center}
\includegraphics[width=0.9\linewidth]{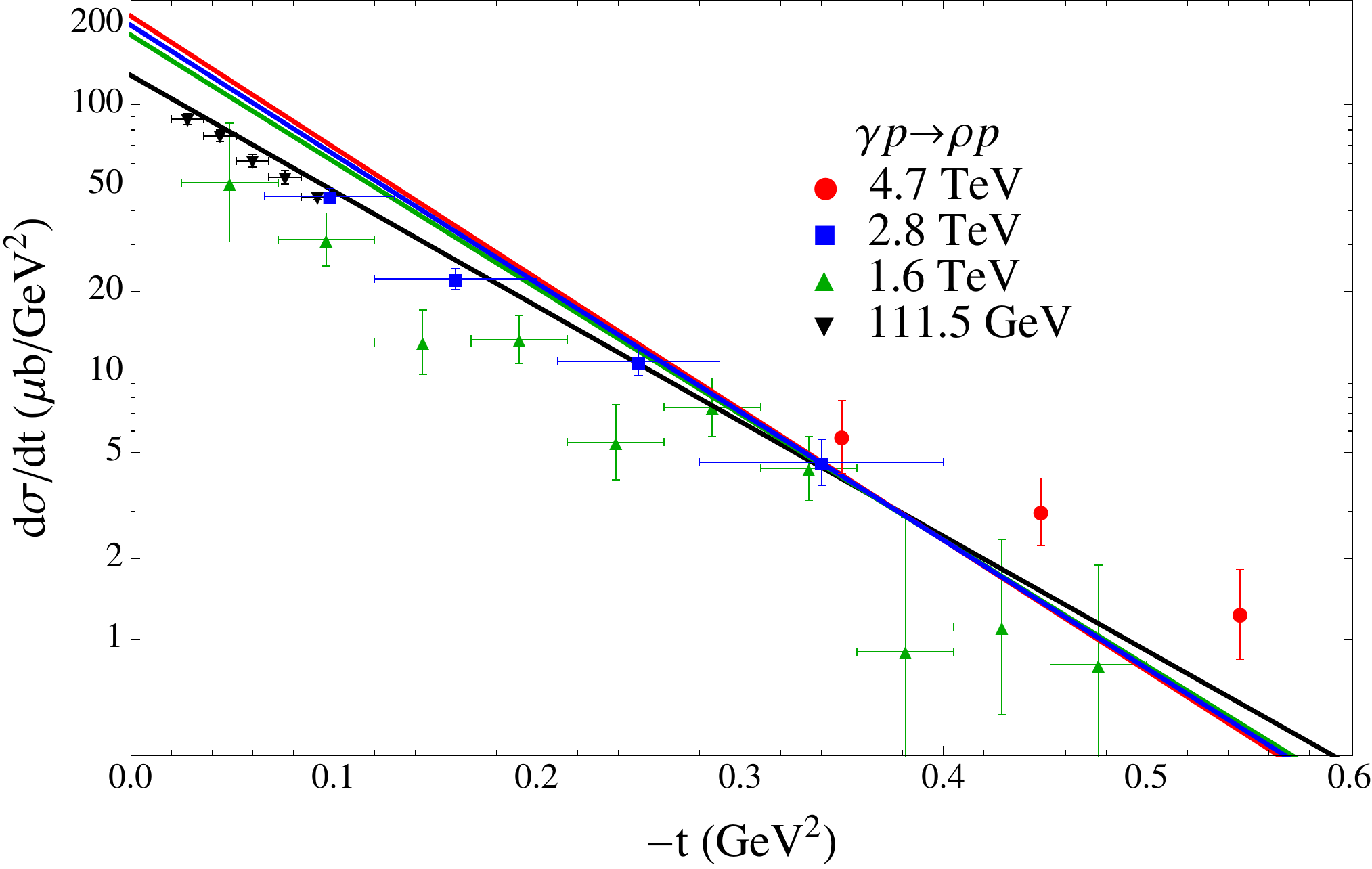}
\includegraphics[width=0.9\linewidth]{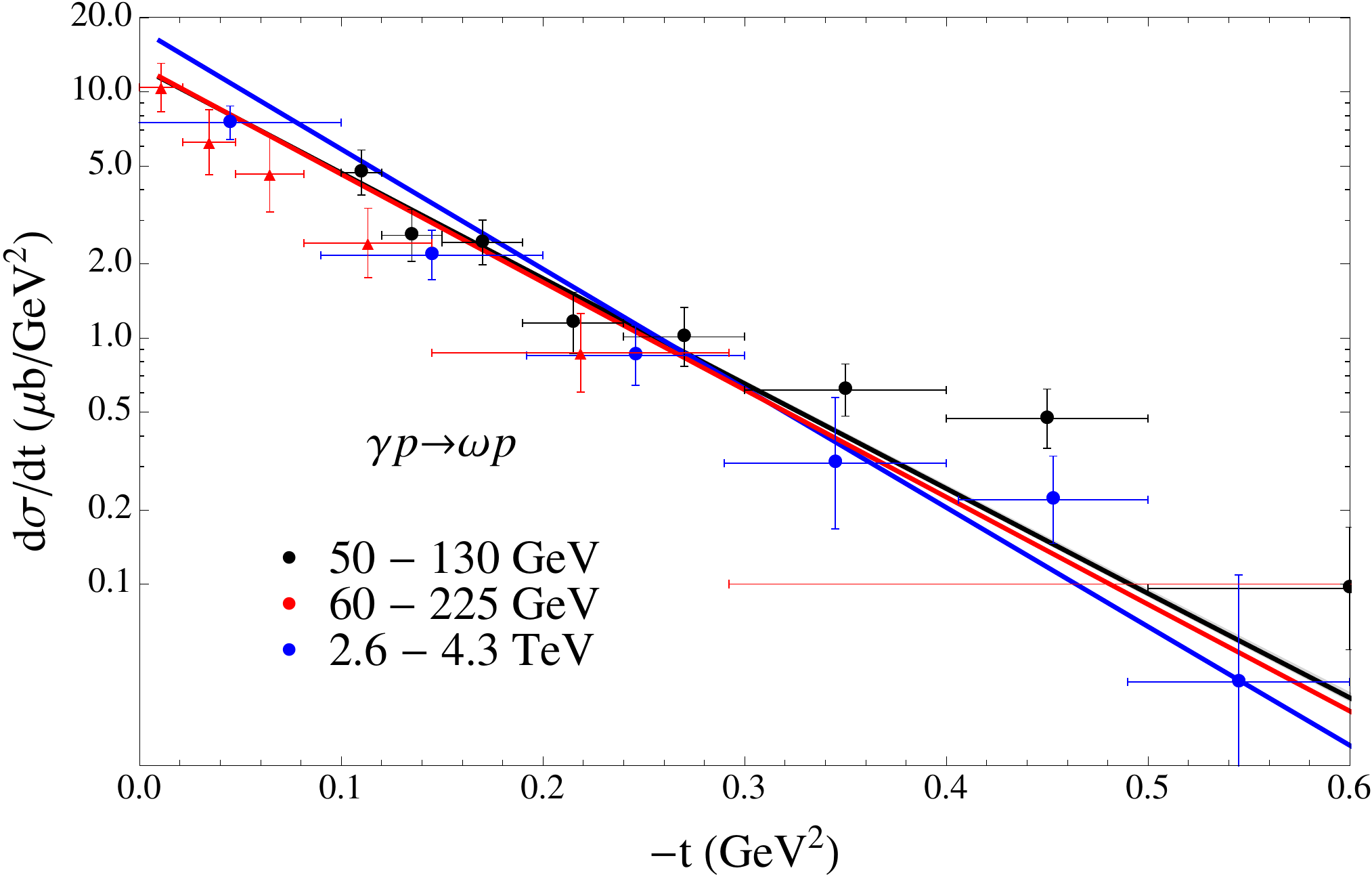}
\end{center}
\caption{\label{fig:dsig-high}$\gamma p \to \rho^0 p$ (top) and  $\gamma p \to \omega p$ (bottom) differential cross sections at high energies. {\it Left panel:} Data from Ref~\cite{Aid:1996bs} (green triangles), Ref~\cite{Chapin:1985mf} (black triangles), Ref~\cite{Breitweg:1999jy} (red circles) and  Ref~\cite{Derrick:1996vw} (blue squares). The higher energies curves overestimate data at low $t$, as expected from the saturation of the unitarity bound. {\it Right panel:} Data from Ref~\cite{Breakstone:1981wk} (black circles), Ref~\cite{Derrick:1996yt} (blue circles) and Ref~\cite{Busenitz:1989gq} (red circles). }
\end{figure}

Our model has been designed to describe the SDMEs, but it is also interesting to compare it with high-energy unpolarized differential cross-section data. We first compare our model to high-energy data in Fig.~\ref{fig:dsig-high}. At energies above $50$ GeV, the Regge exchanges contribute less that 1$\%$ of the differential cross section. The data therefore gives a very good indication of the validity of our Pomeron model. We observe that the overall normalization at $t=0$ is in fairly good agreement with the data. Our phenomenological intercept $\alpha_{\mathbb P} (0)= 0.08$ produces a small rise of the differential cross section in the forward direction. At very high energies, $E_\gamma > 1$ TeV, the data seems to display a slower growth at $t=0$, in agreement with the unitarity bound. However, these energies are far from our region of interest. The $t$-dependence was approximated by a simple exponential falloff, which describes the falloff of the differential cross section in the range $0< -t/m_V^2 \lesssim 1$. We observe deviations from this simple picture at $|t|>0.3$ GeV$^2$.

\begin{figure}[htb]
\begin{center}
\includegraphics[width=0.9\linewidth]{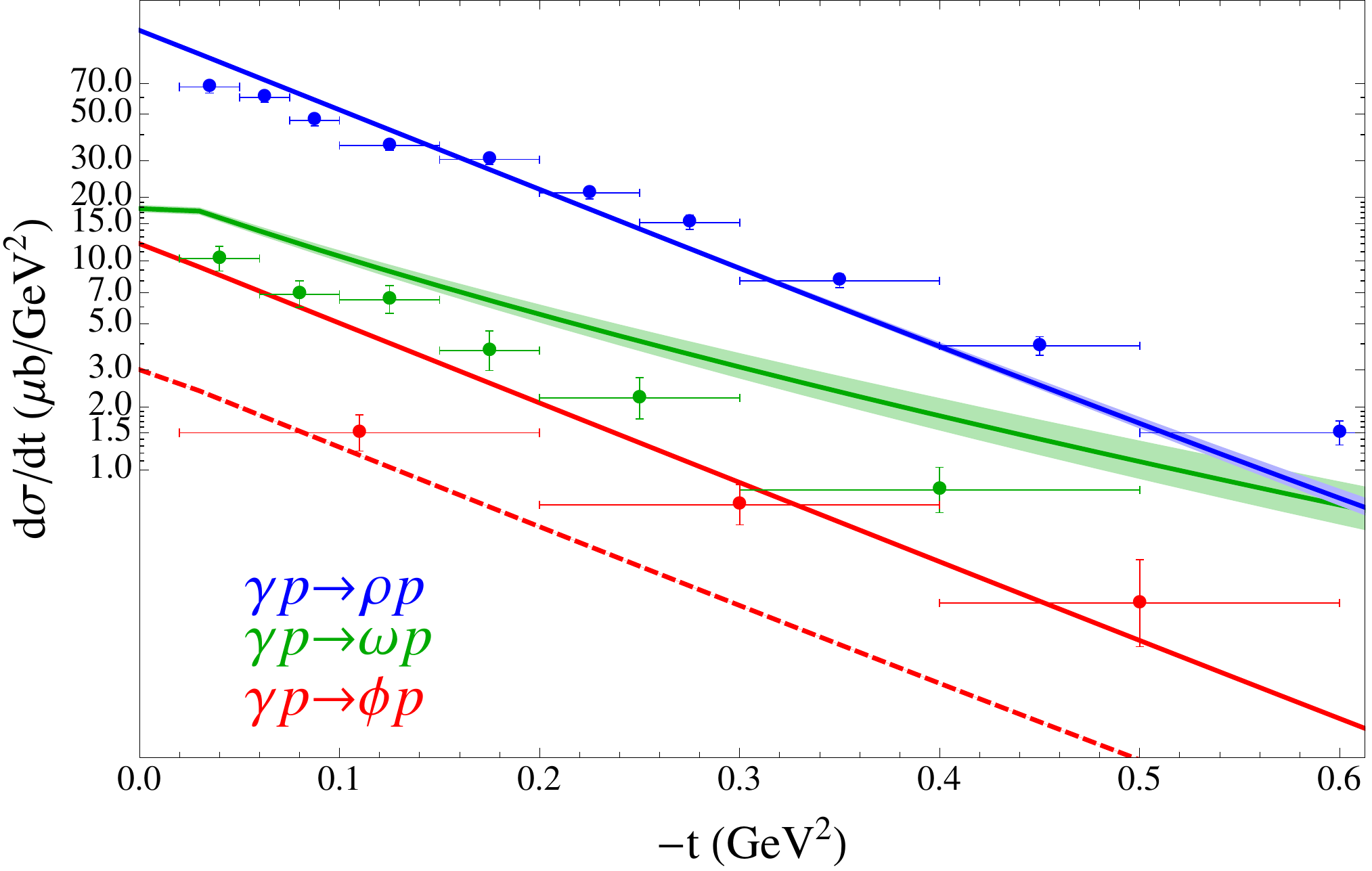}
\end{center}
\caption{\label{fig:dsig-rho-ome-93}$\gamma p \to (\rho^0, \omega,\phi) p$ differential cross section at 9.3 GeV in solid blue, green and red lines respectively. The dashed red line is obtained with a Pomeron coupling reduced by a factor two. The data are taken from Ref~\cite{Ballam:1972eq}.}
\end{figure}

Unfortunately, our model does not compare very well with the $\omega$ and $\phi$ differential cross sections at $9.3$~GeV, as shown in Fig.~\ref{fig:dsig-rho-ome-93}. Although the $\rho^0$ differential cross section is roughly in agreement with our model, the $\phi$ differential cross section is overestimated. We already explained that the leptonic width of the $\phi$ meson led to a Pomeron coupling to $\gamma \phi$ much stronger than the experimental value. This was already observed in the original experimental publication~\cite{Ballam:1972eq}. It has been argued in Ref.~\cite{Barger:1975yw} that the large $\phi$ mass needs to be taken into account. The authors of Refs.~\cite{Behrend:1978ik,Barber:1981fj} corrected the differential cross section by the ratio of the $\phi$ and photon momenta, $(k_\phi/k_\gamma)^2 \approx 0.87$ at $E_\gamma = 9.3$ GeV. This factor is nevertheless not small enough to reproduce the experimental normalization of the $\phi$ differential cross section. As we did for the SDMEs, we reduce the Pomeron coupling $\beta_{0,\phi}^{\mathbb P}$ by a factor of two. The resulting normalization at $t=0$ seems more in agreement with the data.

\section{Conclusions} \label{sec:concl}
We presented a model describing the SDMEs of light vector meson photoproduction. Our model includes $\pi$ and $\eta$ exchanges, whose parameters are fixed. We incorporated the leading natural exchanges: the Pomeron, $f_2$ and $a_2$ exchanges. Their normalizations were determined from the total cross section using the VMD hypothesis. We paid special attention to the $t$-dependence of the various exchanges. We proposed a flexible and intuitive ansatz for the $t$-dependence of each natural exchange. The helicity structure of these exchanges was then inferred from the data on photoproduction of $\omega$ and $\rho^0$ at $E_\gamma = 9.3$ GeV from SLAC. The joint inspection of these two reactions allowed us to assume that the $f_2$ isoscalar exchange must have a small double helicity flip coupling, in addition to a single helicity flip coupling. The $a_2$ isovector exchange was consistent with only a single flip and no double helicity flip coupling. 
 
The model compares well with the nine SDMEs for $\rho^0$, $\omega$ and $\phi$ photoproduction in a wide energy range $E_\gamma \sim 3-9$ GeV, as well as with the unpolarized data in the same energy range. Except for $\rho^0_{1-1}$ in $\omega$ production, the SDME are consistent the factorization of Regge residues.  We made predictions for the future measurements of light meson photoproduction at JLab. Our predictions and our model are available online on the JPAC website~\cite{JPACweb, Mathieu:2016mcy}. With the online version of the model, users have the possibility to vary the model parameters and generate the SDMEs for $\rho^0$, $\omega$ and $\phi$ photoproduction. The code can also be downloaded.

The differential cross section at very high energies, $E_\gamma > 50$~GeV, is well reproduced by our Pomeron exchange. However, the effect of the high-energy approximation led to non negligible deviation in normalization from the data at $E_\gamma = 9.3$ GeV. These deviations appear only in the differential cross section, since they cancel in the ratio of the SDMEs.

\begin{acknowledgments}
We thank C.~Meyer for pointing out the future measurements by the GlueX collaboration.
This work was supported by BMBF,
by the U.S.~Department of Energy under grants No.~DE-AC05-06OR23177 and No.~DE-FG02-87ER40365, by
PAPIIT-DGAPA (UNAM, Mexico) grant No.~IA101717,
CONACYT (Mexico) grant No.~251817, by
Research Foundation -- Flanders (FWO), by
U.S.~National Science Foundation under award numbers PHY-1415459 and PHY-1205019,
and by Ministerio de Econom\'ia y Competitividad (Spain) through grant
No.~FPA2016-77313-P.
\end{acknowledgments}

\appendix
\section{Frames}\label{app:frame}
\begin{figure}[htb]
\begin{center}
\includegraphics[width=.8\linewidth]{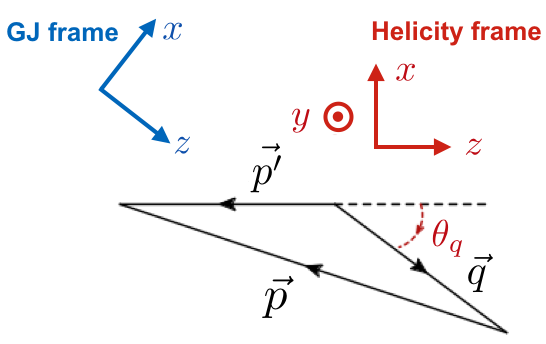}
\\[0.5cm]
\includegraphics[width=.45\linewidth]{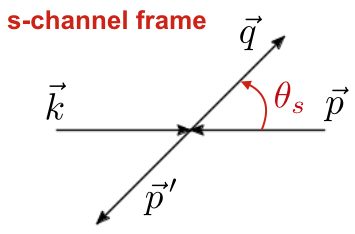}
\includegraphics[width=.45\linewidth]{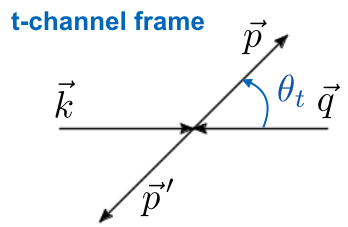}
\end{center}
\caption{\label{fig:frames}Illustration of the frame defined in Appendix~\ref{app:frame}.}
\end{figure}
The properties of helicity amplitudes are best described in two popular frames: the $s$-channel and the $t$-channel frames. The $s$-channel corresponds to the center-of-mass of the reaction $\gamma p \to V p$. The $t$-channel corresponds to the center-of-mass of the reaction $\gamma \bar V \to p \bar p$. These channels are illustrated on Fig.~\ref{fig:frames}.

The angular distribution of a vector meson is analyzed in its rest frame. In the rest frame, the beam, target and recoil form the reaction plane $xz$. The $y$-axis is defined as the cross product between the target and the recoil momenta. For the $z$-axis, the two common choices are the opposite direction of the recoil in the helicity frame, and the beam direction in the GJ frame~\cite{Gottfried:1964nx}.

The helicity amplitudes in these four frames are different. For instance, a boost along the recoil momentum between the $s$-channel and the helicity frames rotates the helicities of the beam, target and recoil. It also transforms the helicity of the vector meson in the $s$-channel into its spin projection along the direction opposite to the recoil in the helicity frame. The summation over beam, target and recoil helicities in the SDMEs is not affected by these rotations. Hence, the SDMEs in the $s$-channel and helicity frames are equivalent. 

Similarly, a boost along the beam direction between the $t$-channel and the GJ frames brings the helicity of the vector in the $t$-channel to its spin projection along the beam direction in the GJ frame. The helicities of the other particles undergo a rotation which does not affect the SDMEs, as demonstrated in Ref~\cite{Gottfried:1964nx}. 

Finally, from the SDMEs in the GJ frame, the SDMEs in the helicity frame are obtained by a rotation of angle $\theta_q$, the angle between the opposite direction of the recoil and the beam direction (see Fig.~\ref{fig:frames})
\begin{align} \label{eq:rot}
\rho_{MM'}|_{GJ} = \sum_{\lambda_V, \lambda'_V} d^1_{M,\lambda_V}(\theta_q)\ \rho_{\lambda_V, \lambda'_V}|_{H}\ d^1_{M',\lambda'_V}(\theta_q),
\end{align}
with $\cos\theta_q = (\beta- \cos \theta_s)/(\beta \cos\theta_s-1)$ and $\beta = \lambda^{1/2}(s,m_p^2,m_V^2)/(s-m_p^2+m_V^2)$. The leading $s$ expression is simply $\cos \theta_q \to (m_V^2+t)/(m_V^2-t)$. 

\section{Spin-Density Matrix Elements} \label{app:sdme}
The relation between SDMEs and helicity amplitudes are well known~\cite{Schilling:1969um}. For completeness, we provide the expressions for the nine SDMEs accessible with a linearly polarized photon beam:
\begin{subequations}
\begin{align}
\rho^0_{00} &= \frac{1}{N}  \sum_{\lambda, \lambda'} 
{\cal M}_{\substack{1, 0 \\ \lambda, \lambda'}} {\cal M}^*_{\substack{1, 0 \\ \lambda, \lambda'}},
\\
\text{Re }\rho^0_{10} &= \frac{1}{2N} \re  \sum_{\lambda, \lambda'} 
\left( {\cal M}_{\substack{1, 1 \\ \lambda, \lambda'}} - {\cal M}_{\substack{1, -1 \\ \lambda, \lambda'}}   \right) {\cal M}^*_{\substack{1, 0 \\ \lambda, \lambda'}}, \\
\rho^0_{1-1} &= \frac{1}{N} \re  \sum_{\lambda, \lambda'} 
{\cal M}_{\substack{1, 1 \\ \lambda, \lambda'}} {\cal M}^*_{\substack{1, -1 \\ \lambda, \lambda'}}, \\
\rho^1_{11} &= \frac{1}{N}  \re \sum_{\lambda, \lambda'}
{\cal M}_{\substack{-1, 1 \\ \lambda, \lambda'}} {\cal M}^*_{\substack{1, 1 \\ \lambda, \lambda'}}, \\
\rho^1_{00} &= \frac{1}{N} \re  \sum_{\lambda, \lambda'} 
{\cal M}_{\substack{-1, 0 \\ \lambda, \lambda'}} {\cal M}^*_{\substack{1, 0 \\ \lambda, \lambda'}},  \\
\rho^1_{1-1} + \im \rho^2_{1-1} & = \frac{1}{N} \sum_{\lambda, \lambda'} 
{\cal M}_{\substack{-1, 1 \\ \lambda, \lambda'}} {\cal M}^*_{\substack{1, -1 \\ \lambda, \lambda'}}, 
\\
\rho^1_{1-1} - \im \rho^2_{1-1} & = \frac{1}{N} \sum_{\lambda, \lambda'} 
{\cal M}_{\substack{1, 1 \\ \lambda, \lambda'}} {\cal M}^*_{\substack{-1, -1 \\ \lambda, \lambda'}},
\\
\re \rho^1_{10} + \im \rho^2_{10} & = \frac{1}{N} \re \sum_{\lambda, \lambda'} 
{\cal M}_{\substack{-1, 1 \\ \lambda, \lambda'}} {\cal M}^*_{\substack{ 1,0 \\ \lambda, \lambda'}}, & 
\\
\re \rho^1_{10} - \im \rho^2_{10} & = \frac{1}{N} \re \sum_{\lambda, \lambda'} 
{\cal M}_{\substack{1, 1 \\ \lambda, \lambda'}} {\cal M}^*_{\substack{-1,0 \\ \lambda, \lambda'}}. & 
\end{align} \label{eq:SDMEs}
\end{subequations}
Of course, the SDMEs and the helicity amplitudes need to be define in the same frame, or in equivalent frames, as explained in the previous section. 
The frame-independent normalization is 
\begin{align}
N & = \frac{1}{2} \sum_{\lambda_\gamma, \lambda_V, \lambda, \lambda'} 
|{\cal M}_{\substack{\lambda_\gamma, \lambda_V \\ \lambda, \lambda'}}|^2.
\end{align}
The implication of helicity conservation at the photon vertex, {\it i.e.}, ${\cal M}_{\substack{\lambda_\gamma, \lambda_V \\ \lambda, \lambda'}} \propto \delta^{\lambda_V}_{\lambda_\gamma} $ can easily be checked in the SDMEs. As can be readily verified with Eqs~\eqref{eq:SDMEs}, this hypothesis leads to vanishing SDMEs except for $\im \rho^1_{1-1}$ and $\im \rho^2_{1-1}$. The SDMEs also provide other useful information concerning the helicity structure of the photon vertex. For instance, the elements $\rho^0_{00}$ and $\rho^0_{1-1}$ give indications about the magnitude of the single-flip contribution and the interference between the nonflip and the double-flip amplitudes. Moreover, they can be used to separate the contributions from natural and unnatural exchanges. Indeed, at high energies, an exchange with positive naturality ($N$) or negative naturality ($U$), satisfies
\begin{align} 
{\cal M}^{\substack{N\\ U}}_{\substack{-\lambda_\gamma, -\lambda_V \\ \lambda, \lambda'}} & = \pm (-1)^{\lambda_\gamma-\lambda_V}
{\cal M}^{\substack{N\\ U}}_{\substack{\lambda_\gamma, \lambda_V \\ \lambda, \lambda'}}.
\end{align}
We can then use six SDMEs to get information about the helicity structure of natural and unnatural components:
\begin{subequations}\label{eq:sdmeNat-unnat}
\begin{align} 
\rho^{\substack{N\\ U}}_{00} &= \frac{1}{2} \left( \rho^{0}_{00} \mp \rho^{1}_{00} \right), \\
%
\text{Re }\rho^{\substack{N\\ U}}_{10} &= \frac{1}{2} \left(  \re \rho^{0}_{10} \mp \re \rho^{1}_{10} \right),  \\
%
\rho^{\substack{N\\ U}}_{1-1} &= \frac{1}{2} \left(  \rho^{1}_{1-1} \pm \rho^{1}_{11} \right). 
\end{align}
\end{subequations}

\section{High-Energy Limit} \label{app:HEL}
At high energies, models for reaction amplitudes simplify.  In this section, we perform the high-energy limit of single-meson exchange interaction and keep the leading-order dependence in $s$, the total energy squared. Our goal is to derive the $t$-dependence arising from the factorization of Regge poles. We consider the reaction $\gamma (k,\lambda_\gamma) p(p,\lambda) \to V(q,\lambda_V) p (p',\lambda')$ in the center-of-mass frame ($s$-channel frame). Let $m_p$ and $m_V$ be the nucleon and vector meson masses, respectively.

\subsection{Unnatural exchanges}
Let us first focus on the pseudoscalar exchanges. According to the factorization theorem for Regge poles, the interaction is a product of a $\gamma V P$ vertex, a Regge factor and a $P NN$ vertex. At the photon vertex we use
\begin{align} \label{eq:pionV}
T_{\lambda_\gamma \lambda_V} & =  -i g_{V P\gamma }\, \varepsilon_{\alpha\beta\mu\nu} \epsilon^\alpha(\lambda_\gamma) \epsilon^{*\beta}(\lambda_V) k^\mu q^\nu.
\end{align}
The polarization vectors, in the $s$-channel, are
\begin{subequations}
\begin{align}
\epsilon^\alpha(\lambda_\gamma) & = \frac{-\lambda_\gamma}{\sqrt{2}} (0,1,\lambda_\gamma i, 0), \\ \nonumber
\epsilon^{*\beta}(\lambda_V) & = \frac{\lambda_V}{\sqrt{2}} (0,-\cos\theta_s,\lambda_V i, \sin\theta_s) \\ & + \frac{1-\lambda_V^2}{m_V}(q_s, E^V_s \sin \theta_s,0,E^V_s \cos\theta_s),
\end{align}
\end{subequations}
where $E_s^V$ and $q_s$ are the energy and momentum of the vector meson in the $s$-channel frame, respectively, and $\theta_s$ is the scattering angle. The expression of the kinematical quantities can be found in the appendix of Ref.~\cite{Mathieu:2015eia}.  
In the center-of-mass frame, the angular dependence of the interaction~\eqref{eq:pionV} is instructive:
\begin{align}
T_{\lambda_\gamma \lambda_V} & \propto \left( \cos \frac{\theta_s}{2} \right)^{|\lambda_\gamma+\lambda_V|} \left( \sin \frac{\theta_s}{2} \right)^{|\lambda_\gamma-\lambda_V|},
\end{align}
with $\theta_s$ the scattering angle in the $s$-channel frame.
This factor, known as the half-angle factor, encodes all the $t$-dependence of the interaction. At large energies, the $t$-dependence of the half-angle factor becomes very intuitive,\footnote{In what follows, we will denote the leading term in $s$ by an arrow.} 
 $\sin\theta_s/2 \to \sqrt{-t/s}$ and $\cos\theta_s/2 \to 1$. Throughout this paper, we neglect the difference between $t$ and $t'$, where $t' = t-t_\text{min}$, since in the kinematical region of interest $t_\text{min}/m_V^2 \to -(m_V/2p_\text{lab})^2$ is on the order of $10^{-3}$ at $p_\text{lab} = 9$ GeV. 

Keeping only the leading term in $s$ of the interaction in Eq.~\eqref{eq:pionV}, we obtain
\begin{align}  \nonumber
T_{\lambda_\gamma \lambda_V} & \to g_{V P\gamma }\frac{m_V^2}{2} \\  \times &
  \left(
\lambda_\gamma \delta_{\lambda_V,\lambda_\gamma}
- \sqrt{2} \frac{\sqrt{-t}}{m_V} \delta_{\lambda_V,0} + \frac{-t}{m_V^2} \lambda_\gamma\delta_{\lambda_V,-\lambda_\gamma} \right).
\label{eq:pionV2}
\end{align}
This example illustrates a general statement: each helicity flip ``costs" a factor of $\sqrt{-t}/m_V$. The mass scale associated to the factor $\sqrt{-t}$ can only be $m_V$. For completeness, we derive the decay width from the interaction~\eqref{eq:pionV}:
\begin{align} \label{eq:widthpion}
\Gamma(V\to \gamma P) & = \frac{g_{V P\gamma }^2}{96\pi} \left( \frac{m_V^2-m_P^2}{m_V} \right)^3. 
\end{align}
We use Eq.~\eqref{eq:widthpion} to extract the couplings from the decay widths. The relevant couplings are summarized in Table~\ref{tab:pi}.
\begin{table}[htb]
\centering
\caption{Vector meson radiative decay widths and pseudoscalar exchange couplings.
\label{tab:pi}}
\begin{tabular}{c|cc|cc}
 $V$ &$\Gamma(V\to \gamma \pi^0)$ & $g_{V\pi\gamma} $ & $\Gamma(V\to \gamma \eta)$ & $g_{V\eta\gamma} $  \\
\hline
$\omega$ & 703 keV &  0.696 GeV$^{-1}$ & 44.8 KeV &  0.479 GeV$^{-1}$\\
$\rho^0$ & 89.6 keV &  0.252 GeV$^{-1}$ & 3.91 KeV &  0.136 GeV$^{-1}$\\
$\phi$ & 5.41 keV &  0.040 GeV$^{-1}$ & 56.8 KeV &  0.210 GeV$^{-1}$\\
\end{tabular}
\end{table}

The considerations at the photon vertex apply equally well at the nucleon vertex. For an unnatural spin-zero exchange, there is only one possible structure at the nucleon vertex:
\begin{align}
g_{P NN}\bar u (p',\lambda) \gamma_5 u (p,\lambda) &\to g_{P NN} \sqrt{-t} \delta_{\lambda',-\lambda}.
\end{align}
There is one unit of helicity flip associated with the factor $\sqrt{-t}$. In this case the scale factor (nucleon mass) is implicitly removed by our spinor normalization $\bar u(p,\lambda) u(p,\lambda) = 2m$. For the $\pi$-nucleon and $\eta$-nucleon couplings, we take $g^2_{\pi NN}/4\pi = 14$ 
\cite{Timmermans:1990tz,Baru:2011bw,Baru:2010xn,Arndt:2006bf,Ericson:2000md,Arndt:1994bu,Arndt:1990cn,Perez:2016aol}, and $g^2_{\eta NN}/4\pi = 0.4$ is the value we used in our fixed-$t$ dispersion relation analysis of $\eta$ photoproduction~\cite{Nys:2016vjz}
 based on the available literature~\cite{Tiator:1994et,FernandezRamirez:2007fg,Dumbrajs:1983jd,Neumeier:2000fb,Kirchbach:1996kw,Saghai:2001yd}.

The couplings we determined are normalized at the pseudoscalar pole. We then add a factor $\pi\alpha'/2$ to the Regge factor in Eq.~\eqref{eq:ReggePi} such that
\begin{align}
\lim_{t\to m_P} (t-m_P)\frac{\pi \alpha_P'}{2} R^P(s,t) =1.
\end{align}
The Regge trajectory is $\alpha_P(t) = \alpha_P'(t-m_\pi^2)$ with $\alpha_P' = 0.7$ GeV$^{-2}$. We choose the same trajectory for both $\pi$ and $\eta$ exchange. As explained in Sec.~\ref{sec:comp}, this enhances the $\eta$ pole to compensate for the Pomeron normalization in the $\phi$ photoproduction SDMEs. 
Finally, collecting all the pieces, we arrive to the amplitude in Eq.~\eqref{eq:pion} for a $\pi$ or $\eta$ exchange in the high-energy limit with the normalization $\beta^{P}_{0,V} = (1/4)\pi \alpha' m_V^2 g_{VP\gamma} g_{PNN}$.

It is instructive to derive the SDMEs for only a $\pi$ exchange in both the GJ and helicity frames. The SDMEs induced by a $\pi$ exchange take a simple form in the GJ frame, {\it i.e.}, all SDMEs are zero except for $\rho^1_{1-1} = - \im \rho^2_{1-1} = -\frac{1}{2}$. This is of course expected since the $\pi$ in its rest frame only has the spin projection zero. We can easily get the SDMEs for a $\pi$ exchange in the helicity frame from the rotation in Eq.~\eqref{eq:rot}: 
\begin{subequations} \label{eq:sdmepion}
\begin{align}
\rho^0_{00} & = \rho^1_{00} = \frac{-2 t /m_V^2}{\left(1-t/m_V^2 \right)^2}, \\
\rho^0_{1-1} & = -\rho^1_{11} = \frac{-t /m_V^2}{\left(1-t/m_V^2 \right)^2}, \\
\re \rho^0_{10} & = \re \rho^1_{10} = \frac{-1}{\sqrt{2}} \frac{\sqrt{-t}}{m_V} \frac{1+ t /m_V^2}{\left(1-t/m_V^2 \right)^2}, \\
\rho^0_{11} & = -\frac{1}{2} \frac{1+ t /m_V^2}{\left(1-t/m_V^2 \right)^2}, \\
\im \rho^2_{1-1} & = -\frac{1}{2} \frac{1- t /m_V^2}{\left(1-t/m_V^2 \right)^2}, \\
\im \rho^2_{10} & = \frac{-1}{\sqrt{2}} \frac{\sqrt{-t}}{m_V} \frac{1}{\left(1-t/m_V^2 \right)^2}.
\end{align}
\end{subequations}
In the case of a single exchange, the SDMEs depend only on the details of the photon vertex. The only scale that arises is the mass of the vector meson.

\subsection{Natural exchanges}
The two guiding rules, the factorization of Regge poles and the factor of $\sqrt{-t}$ for each unit of helicity flip, equally apply to natural exchanges. We can then postulate the general form in Eq~\eqref{eq:factor}. Since the use of effective Lagrangians is very popular, it is instructive to compare our model in Eqs~\eqref{eq:topVnat} and \eqref{eq:bottomVnat} to these types of interactions. 

Let us start with the standard interaction for a Pomeron exchange~\cite{Oh:2000zi, Lesniak:2003gf}
\begin{align} \nonumber
{\cal M}^{\mathbb P}_{\substack{\lambda_\gamma, \lambda_V \\ \lambda, \lambda'}}(s,t) & = \epsilon^*_\nu(q,\lambda_V)  \left[k^\mu \ 
\epsilon^\nu(k,\lambda_\gamma)  
 - k^\nu  \epsilon^\mu(k,\lambda_\gamma)  \right] \\
& \times  \bar u(p',\lambda')\gamma_\mu  u(p,\lambda).
 \label{eq:pom}
\end{align}
At leading order in $s$, we have $k^\mu \to n_+^\mu \sqrt{s}/2 $, with $n_\pm^\mu = (1,0,0,\pm 1)$, and the helicity structure at the nucleon vertex is simply 
\begin{align} \label{eq:nucleon1}
\bar u(p',\lambda')\gamma^\mu  u(p,\lambda)\to \sqrt{s}\, \delta_{\lambda,\lambda'}\, n_-^\mu.
\end{align} 
Only the first term in the bracket in Eq.~\eqref{eq:pom} survives. From the result
\begin{align} \label{eq:struc1}
\epsilon^*(q,\lambda_V)\cdot \epsilon(k,\lambda_\gamma)  & \to - \delta_{\lambda_V,\lambda_\gamma} + \delta_{\lambda_V,0} \frac{\lambda_\gamma}{\sqrt{2}} \frac{\sqrt{-t}}{m_V},
\end{align}
we conclude that this model for the Pomeron implicitly includes a single helicity flip structure and is not therefore purely helicity conserving. A more flexible model can be obtained with more general interactions. In order to determine all of the possible structures at both the photon and the nucleon vertices, let us first observe that in a factorizable model, the top and bottom vertices are linked by a propagator transverse to the momentum transferred. The propagator removes the $x$ component since $(q-k)^\mu \to (0,\sqrt{-t},0,0)$ at leading order in $s$. Secondly, the general structures at the nucleon vertex are easily obtained. In addition to Eq.~\eqref{eq:nucleon1}, we can have an nucleon helicity-flip interaction
\begin{align}\label{eq:nucleon2}
\bar u(p',\lambda') \left(2p^\mu- \gamma^\mu \right) u (p,\lambda) &\to 2 \lambda\sqrt{-t} \delta_{\lambda',-\lambda}\, n_-^\mu.
\end{align}
Note that any $p'$ momentum can be substituted by $p$ since the difference is orthogonal to the propagator. We summarized the two possible structures at the nucleon vertex in Eqs~ \eqref{eq:nucleon1} and \eqref{eq:nucleon2} in Eq.~\eqref{eq:bottomVnat}. 

At the photon vertex, the only tensorial structures that connect to the nucleon vertex and survive at leading order in $s$ are $k^\mu \to n_+^\mu \sqrt{s}/2 $  and $\epsilon^*( q,\lambda_V) \to (1-\lambda^2_V) (q/m_V) n_+^\mu$. We can then form a single helicity-flip coupling at the photon vertex with the interaction
\begin{align} \label{eq:struc2}
\epsilon^{\mu *}( q,\lambda_V)\ q\cdot \epsilon(k,\lambda_\gamma) & \to \frac{1}{2} \frac{\lambda_\gamma}{\sqrt{2}}\frac{\sqrt{-t}}{m_V} \delta_{\lambda_V,0}\, n_+^\mu.
\end{align}
Finally, since the maximum helicity difference between a photon and a vector meson in their center of mass is two, a tensor exchange should involve all possible relevant structures at the photon vertex. Indeed, we find that a double-flip structure can arise with the interaction between a photon, vector and tensor~\cite{Levy:1975fk}:
\begin{align}\label{eq:struc3}
\epsilon^*(q,\lambda_V)\cdot k\ \epsilon(k,\lambda_\gamma) \cdot q \to \lambda_\gamma \lambda_V \frac{t}{2}.
\end{align}
We then conclude that, in addition to the nonflip interaction in Eq.~\eqref{eq:struc1}, the general structure with a photon, vector and natural exchange also includes the single-flip interaction in Eq.~\eqref{eq:struc2} and the double-flip interaction in Eq.~\eqref{eq:struc3}. To leading order in $s$, we summarize these interactions with the intuitive vertex in Eq.~\eqref{eq:topVnat}. 

\medskip
In our model we added a helicity-independent exponential factor $b_N$ to reproduce the energy-independent shrinkage of the differential cross section. 
This feature is generally described by exponential factors~\cite{Schilling:1968awz, Gotsman:1970dy, Ader:1970sg, Barker:1973dm}, gamma functions~\cite{Irving:1977ea,Laget:2000gj} or dipole form factors~\cite{Donnachie:1983hf,Laget:1994ba, Oh:2000zi,Sibirtsev:2003qh,Yu:2017vvp}. This $t$-dependence originates from the energy dependence of the nearest cross-channel singularity. For the given Regge exchanges, these are the $f_2(1270)$ and $a_2(1320)$ tensor mesons. The energy dependence of these singularities in the cross channel can be described by Breit-Wigner line shape in $t$, the relevant energy variable in the cross channel:
\begin{align}
BW_E(t) & = \frac{m_E \Gamma_E}{m_E^2-t-i m_E \Gamma_E},
\end{align}
where $m_E$ and $\Gamma_E$ are the mass and the width of the $f_2$ and $a_2$ tensor mesons. Its effect in the physical region of the direct channel can be modeled by a simple exponential falloff, \textit{i.e.}, $|BW_E(t)|^2 \approx  |BW_E(0)|^2 e^{2b_E t}$ for $t \in [-m_\omega^2, 0]$. We determine $b_E$ at $t = - m_\omega^2/2$, the middle point of the interval $t \in [-m_\omega^2, 0]$. We find $b_{f_2} = 0.55$ GeV$^{-2}$ and $b_{a_2} = 0.53$ GeV$^{-2}$. 

The $t$-dependence of the Pomeron is often described by the following dipole form factors~\cite{Donnachie:1983hf,Laget:1994ba, Oh:2000zi,Sibirtsev:2003qh,Yu:2017vvp}
\begin{subequations} \label{eq:FF}
\begin{align}
F_1(t) & = \frac{4m_p^2-2.8 t}{(4m_p^2-t)(1-t/t_0)^2} ,\\
F_V(t) & = \frac{1}{1-t/m_V^2} \frac{2\mu_0^2+ m_V^2}{2\mu_0^2+ m_V^2-t},
\end{align}
\end{subequations}
with $\mu_0^2 = 1.1$ GeV$^2$ and $t_0=0.7$ GeV$^2$. The form factor $F_1(t)$ is the dipole approximation of the nucleon Dirac form factor~\cite{Donnachie:1983hf}, and $F_V(t)$ is an empirical form factor at the photon vertex.\footnote{We have chosen the normalization such that $F_V(0) = 1$.} As for the Regge exchanges, we approximate this form factor by an exponential falloff at $t_0 = -m_\omega^2/2$, $F_1(t_0)F_V(t_0) =  e^{b_{\mathbb P} t_0}$. Under this approximation, we obtain $b_{\mathbb P} = 3.60$ GeV$^{-2}$. 

\bibliographystyle{apsrev4-1}
\bibliography{quattro}

\end{document}